# Controlled Chemical Signaling between Enzymatic Nanomotors

Shuqin Chen[1,2], Giorgio Lovato[3], Oriol Jutglar Soler[1], Daniel Sánchez-deAlcázar[1], Ramin Golestanian[3,4*], Samuel Sánchez[1,5*]

[1]Institute for Bioengineering of Catalonia (IBEC), The Barcelona Institute for Science and Technology (BIST), Baldiri i Reixac 10-12, Barcelona 08028, Spain
[2]Department of Physics, University of Barcelona, Martí i Franquès 1, Barcelona 08028, Spain
[3]Max Planck Institute for Dynamics and Self-Organization (MPI-DS), D-37077 Göttingen, Germany
[4]Rudolf Peierls Centre for Theoretical Physics, University of Oxford, Oxford OX1 3PU, United Kingdom
[5]Catalan Institute for Research and Advanced Studies (ICREA), Passeig Lluís Companys 23, Barcelona 08010, Spain

Corresponding authors:
*Ramin Golestanian: ramin.golestanian@ds.mpg.de
*Samuel Sánchez: ssanchez@ibecbarcelona.eu

## Abstract

The coordinated interactions between organisms enhance collective functionality, a feature that artificial systems such as enzymatic nanomotors seek to replicate. A key objective, yet still a major challenge, is to achieve chemical communication among nanomotors. Progress has been limited by the difficulties in verifying effective signaling processes, including chemical signal propagation and the response of receiving nanomotors. Here, we address this challenge using an enzymatic nanomotor system that demonstrates communication between two populations through generically non-reciprocal phoretic response. A primary swarm of glucose-responsive nanomotors migrates toward a glucose gradient while producing $H_2O_2$ as a diffusible communication signal. This self-generated chemical gradient then acts as a chemoattractant for a secondary swarm of catalase-powered nanomotors. Through carefully designed experiments, we visualize the propagating $H_2O_2$ gradient and quantify the spatiotemporal response of the receiver nanomotors to the chemical front. Combined experimental and theoretical analysis has revealed that the synergy between different combinations of chemo-attractive and chemo-repulsive mobilities and catalytic rates of consumption and production of substrates and products gives rise to a wealth of different collective responses in the system. This work represents a step toward programmable synthetic systems at the collective level, broadening the functionality of chemical nanomotors and opening opportunities for future hybrid living-synthetic systems.

## Introduction

In biological systems, long-range interactions enable functional units such as protein-complexes, organelles, cells, and organisms to perform complex tasks in coordination[1,2]. In particular, chemotaxis---the ability to sense and respond to chemical gradients via a variety of different mechanisms---provides a ubiquitous medium for programmable biological function, ranging from single- and collective- cell dynamics to morphogenesis and immune response. The collective intelligence achieved in such chemotactic mechanisms arises from individual agents responding to local cues rather than centralized control, a fascinating strategy that synthetic systems seek to emulate and harness for biomedical applications[3–5]. However, engineering artificial systems that mimic such complex biological behaviors and exhibit these self-organized, signal-driven interactions remains challenging[6]. Enzymatic nanomotors (ENMs), which exhibit non-equilibrium dynamics associated with the catalytic conversion of chemical fuels[7], offer a promising platform for realizing such cooperative behaviors without direct interparticle contact or external guidance[4], provided they can be shown to have the capacity for a level of programmable complexity that rivals what is observe in living systems.

A particular characteristic of the chemotactic activity of enzymatic nanomotors is their capacity to exhibit non-reciprocal interactions[8]. These interactions, which have so far been experimentally demonstrated for emulsion droplets[9] and catalytic colloids[10] but not for enzymes or ENMs, can be engineered to lead to a wealth of complex behaviors[11,12], including curation of metabolic networks relevant to the early stages of life formation[13]. Based on its catalytic function, an ENM produces or consumes certain chemicals with rates that we call *activities*, which derive from its enzymatic function, and responds to those chemicals and others that are produced or consumed by other ENMs with *mobilities*, which derive from its surface chemistry and can be tailored[8]. The engineering challenge is to identify the right combinations of ENMs with activities and mobilities that can acquire different signs and demonstrate how to control their interactions.

Previous studies have experimentally demonstrated chemotaxis in enzyme-powered nanomotors, including systems driven by urease[14,15], catalase[16], glucose oxidase[17], ATPase[18], and acid phosphatase[19]. Mechanistic studies have shown that a combination of phoretic response and binding-induced changes in the diffusion coefficient can determine whether an enzyme exhibits a positive or negative effective mobility[18,20,21]. Notwithstanding these studies, systematic investigation and quantification of communication-enabled coordinated behavior between two populations of nanomotors remains very limited. Here, we present the coordinated behavior of two populations of enzymatic nanomotors, based on glucose oxidase (GOxNMs) and catalase (CatNMs), mediated by self-sustained chemical gradients and non-reciprocal phoretic interactions. Glucose, a central biological energy source, is often distributed in gradients across cell membranes, vasculature, and tissues[22], whereas $H_2O_2$ serves as a key signaling molecule that regulates processes such as cell proliferation, differentiation, and migration[23]. Their bioavailability makes both fuels suitable candidates for studying biofuel-mediated interactions between living and synthetic systems. Conventional platforms often rely on steep chemical gradients generated between adjacent channels or opposing reservoirs with one side filled with fuel sources[24,25], which may introduce unwanted flows and ENM cross-migration that interfere with experimental observations. To overcome this limitation, we designed a microfluidic setup that establishes a stable, diffusion-controlled fuel gradient. The system consists of a three-channel microfluidic chip with an agarose gel-filled middle channel confined to a height of 70 μm, which acts as a gate that enables the steady fuel release and maintains a stable gradient across the channel. We demonstrated the collective motion of GOxNMs towards glucose gradients and CatNMs towards $H_2O_2$ gradients. The ENM distributions under chemical gradients were imaged using the green fluorescence of fluorescein isothiocyanate (FITC), which was conjugated to the ENM chassis during synthesis. We then examined the communication signaling between the two ENM populations: a glucose gradient first recruits GOxNMs, which release $H_2O_2$ and thereby form a self-sustained gradient that recruits a secondary swarm of CatNMs. We propose this strategy as an approach to achieve coordinated behavior between two ENM populations mediated entirely by chemical signals. From a comprehensive theoretical analysis of the experimental observations and measurements, we conclude that GOxNMs are attracted to their substrate (glucose) and repelled from their product ($H_2O_2$), whereas CatNMs are attracted to their substrate ($H_2O_2$) and product ($O_2$). Moreover, CatNMs are repelled by the substrate of GOxNMs (glucose). Since the product of GOxNMs ($H_2O_2$) is the substrate for CatNMs the pattern of activities and mobilities for the two ENMs leads to an effective non-reciprocal interaction between the two ENMs.

## Results and discussion

**Establishment of chemical gradients.** We used a three-channel microfluidic chip in which the middle channel was filled with agarose gel to serve as a gate for the stable release of fuel and enabled the formation of a static and sustained gradient across the channels. The detailed experimental setup is illustrated in Fig. 1**a** and described in the Methods. We examined the establishment of gradients by imaging the diffusion of fluorescent Rhodamine B (RhB). As shown in Fig. 1**b**, the RhB solution gradually diffused through the middle channel over 20 min, forming a gradient from high concentration in the left to lower concentration in the right channel. RhB diffusion in the right channel is shown in Figs. 1**c-e**. Over 30 min, RhB at different concentrations diffuses into the right channel through the central gel. The normalized fluorescence intensity profiles (Figs. 1**f-h**), extracted from the dashed lines in Figs. 1**c-e**, correspond to this diffusion process. The results reveal that fuel diffusion increases with fuel concentration. Furthermore, the addition of nanomotors into the right channel accelerates fuel diffusion, likely due to enhanced fluid mixture. We also quantified the diffusion of $H_2O_2$ into the right channel (Fig. S1, Supporting Information). When different concentrations of $H_2O_2$ were introduced into the left channel, the overall $H_2O_2$ concentration in the right channel increased gradually over time, as measured every 10-minute interval.

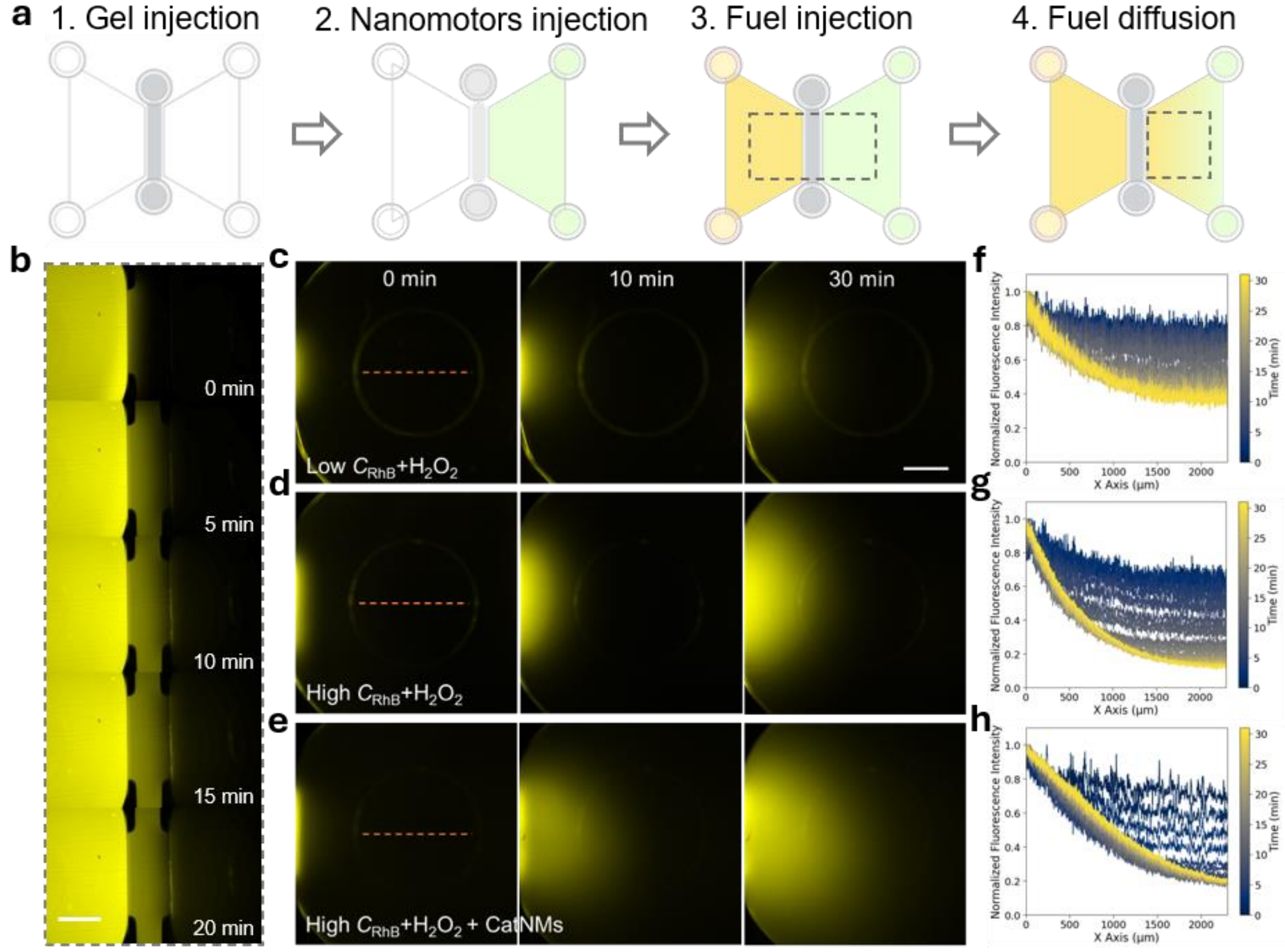

**Fig. 1 The generation of a fuel gradient for ENMs. a** Schematic of the experimental workflow for generating a fuel gradient. The dashed square in step 3 marks the observation area shown in panel **b**, and the dashed square in step 4 indicates the imaging area for panels **c-e**. **b** A time sequence of images show the diffusion of RhB across a three-channel microfluidic chip. Scale bar: 1 mm. **c-e** Time series of images showing RhB diffusion in the right channel across different conditions: **c** Low concentration of RhB ($C_{RhB}$, 2 mM) mixed with $H_2O_2$ (25 mM) diffuses into PBS solution; **d** High $C_{RhB}$ (8 mM) mixed with $H_2O_2$ diffuses into PBS solution; **e** High $C_{RhB}$ (8 mM) mixed with $H_2O_2$ diffuses into a channel filled with CatNMs. Scale bar: 1 mm. **f-h** Line profile of normalized fluorescence intensity analyzed from the dashed lines in **c-e**, respectively.

**Collective behavior of GOxNMs under glucose gradients.** To minimize the initial drift effect of nanomotor injections, we kept the chip still for 20 min after introducing the nanomotors. Fuel was introduced into the left channel, and the motion of nanomotors was analyzed in both the near-fuel region ($P_1$) and the far-fuel region ($P_2$), (Fig. 2**a**, see Methods). We performed single-particle tracking for GOxNMs in glucose gradients (Figs. 2**b-i**). The $x$ and $y$ components of the motion were analyzed separately, revealing purely diffusive behavior in the $y$ direction, which is orthogonal to the fuel front. Consistently, the velocity towards the externally imposed fuel gradient reached plateau after approximately 20 min (Figs. 2**b**). The obtained effective diffusion coefficients (Fig. 2**c**) remain close in magnitude for both positions, while staying strictly below the single colloid Einstein–Stokes relation of around 1.01 $\mu m^2 s^{-1}$. This reduction likely reflects the formation of small colloidal clusters, which diffuse more slowly than individual particles. Over time we observe a consistent upward trend of the diffusion coefficient. This relative enhancement of diffusion can be attributed to the fact that over time the global fuel and product concentrations inside the chamber rise, increasing the activity of the enzymes linked on the particle surface. Cluster formation kinetics (monomers, dimers, etc) under changing chemical concentrations, and consequently, the effective colloidal interactions, can explain this behavior[26].

Particle trajectories and orientations within 5 s were recorded and shown in Figs. 2**d**-**i**. In the near-fuel region $P_1$, GOxNMs exhibit consistent trajectories toward the fuel throughout the observation durations (Fig. 2**d**, **e**). Trajectory directionality analysis (Fig. 2**f**), in which the initial position of each nanomotor was set as the origin and the 5 s displacement was mapped across 24 radial sectors, showed that the majority of nanomotor trajectories pointed towards the 90º-270º area. However, in the far-fuel region $P_2$, both the trajectories (Figs. 2**g**, **h**) and the orientations display random motion in all directions over 30 min, revealing that the glucose front has yet not reached that region.

At the collective level, the ENM distributions were imaged through the green fluorescence of FITC. Three glucose gradient conditions (with 0, 25, and 100 mM concentration difference across the chamber) were tested. Under no-gradient conditions (0.1× phosphate-buffered saline, PBS), both bright-field and fluorescence images at 0 min and 30 min reveal a homogeneous distribution of GOxNMs (Fig. 2**j**). In contrast, introduction of a glucose gradient with 25 mM concentration difference promoted GOxNMs accumulation on the gel side (Fig. 2**k**), an effect that became more pronounced at 100 mM glucose, as evidenced by increased fluorescence intensity on the gel-proximal side (Fig. 2**l**, video S1). Fluorescence intensity profiles along the x-axis were analyzed in the selected regions in Figs. 2**j**-**l**, positioned 200 μm away from the gel interface to minimize boundary effects, demonstrating distinct distribution patterns across different conditions. In the absence of a gradient (Fig. 2**m**), normalized mean intensity remained identical along the x-axis with minimal fluctuations. However, under glucose gradient with 25 mM concentration difference, the gel-adjacent regions exhibited a progressive increase in fluorescence intensity over time (Fig. 2**n**), confirming directed collective migration toward higher glucose concentrations. This heterogeneity became even more pronounced at 100 mM glucose concentration difference (Fig. 2**o**). Fluorescence centroid displacement analysis was also performed (Fig. S2), revealing collective nanomotor migration. At 100 mM glucose concentration difference, the horizontal fluorescence centroid displacement increases fourfold compared with nanomotors under no-gradient conditions.

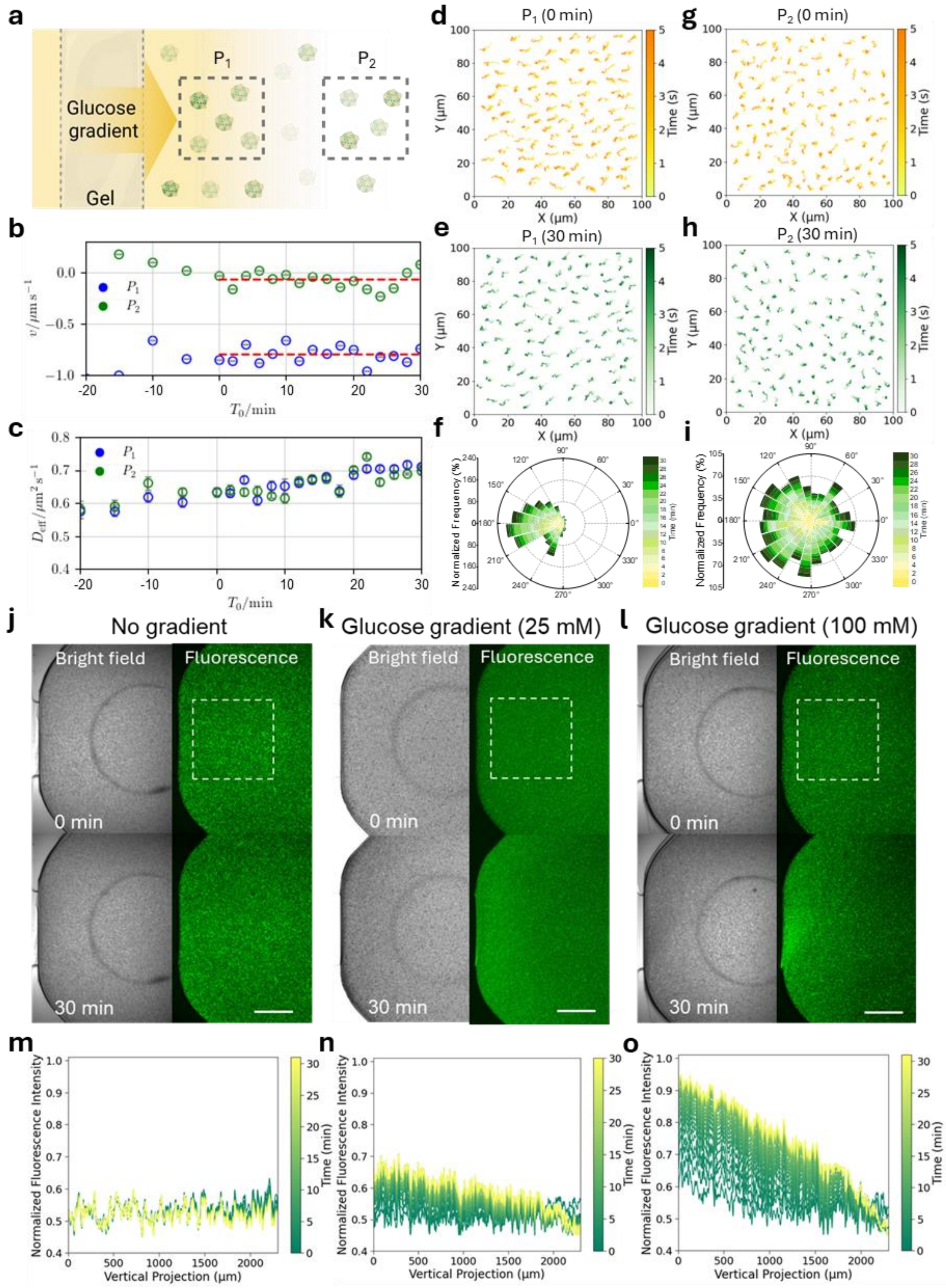

**Fig. 2 Single-particle tracking and collective behavior of GOxNMs in glucose gradients. a** Schematic of GOxNMs under a glucose gradient. Dashed squares $P_1$ and $P_2$ are the measurement positions for single-particle motion analysis. **b** The velocities of particles at $P_1$ (blue) and $P_2$ (green) are shown from the moment fuel is introduced. Over a period of 20 minutes, a clear relaxation of the velocities toward a plateau is visible (from -20 to 0 minute). The red dashed lines represent the average velocity from 0 to 30 minutes. Data are from N = 10 independent experiments, with approximately 100 particles analyzed per experiment. **c** Effective diffusion coefficients are roughly equal in the two measurement positions and increase slightly over time. **d** Representative 5-second trajectories of particles at 0 min and **e** 30 min at $P_1$. **f** Polar map of nanomotor orientations within 5 s at 2-minute intervals at $P_1$. Approximately 100 particles were analyzed. **g** Representative 5-second trajectories of particles at 0 min and **h** 30 min at $P_2$. **i** Polar map of nanomotor orientations within 5 s at 2-minute intervals at $P_2$. Approximately 100 particles were analyzed. **j-l** Representative bright-field and fluorescence images of GOxNMs filled in the right channel of the chip under different conditions: **j** no gradient (0.1× PBS solution), **k** glucose gradient (25 mM concentration difference), and **l** glucose gradient (100 mM concentration difference), captured at 0 and 30 minutes. Scale bar: 1 mm. **m-o** Normalized fluorescence intensity profiles derived from the dashed squares in **j-l**, respectively.

**Collective motion of CatNMs under $H_2O_2$ gradients.** We next performed single-particle tracking for CatNMs in $H_2O_2$ gradients (Fig. 3**a**). As in the glucose-GOxNMs condition, the velocity towards the $H_2O_2$ fuel gradient reaches a plateau value after approximately 20 min (Figs. 3**b**), and the effective diffusion coefficients at $P_1$ and $P_2$ display a consistent upward trend and remain close in magnitude (Fig. 3**c**). However, the trajectory and orientation analyses reveal that CatNMs at both $P_1$ and $P_2$ move toward the fuel side. In $P_1$, particle trajectories (Figs. 3**d**, **e**) and particle orientations (Fig. 3**f**) showed a clearly biased motion, with the particle orientation falling within the 150º-210º area. In $P_2$, particle trajectories and orientations are time dependent. Initially, the motion appears random (Fig. 3**g**), but over the course of 30 min, the particles progressively reoriented toward the fuel side (Fig. 3**h**). Figure 3**i** clearly illustrates this evolution, showing that over 30 min, the majority of nanomotors fall within the 90º-270º area. A deeper polar map analysis of the first 4 min reveals irregular motion in all directions, with 40.6% of nanomotors oriented leftward and 59.4% rightward (Fig. 3**j**). From 6 min onward, orientation toward the fuel source is more prevalent, with 75.8% nanomotors aligning toward the left side (Fig. 3**k**).

The collective behavior of CatNMs was then examined under three conditions. Bright-field and fluorescence images (Figs. 3**l**-**n**, video S2) show the distribution of CatNMs captured at 0 min and 30 min. Compared to the no-gradient conditions (Fig. 3**l**), under an $H_2O_2$ gradient with 25 mM concentration difference (Fig. 3**m**), the CatNMs reacted with the fuel and produced oxygen bubbles that grow larger near the fuel side. This region also exhibited an increased fluorescence intensity, as confirmed by the normalized fluorescence intensity distribution (Figs. 3**o**, **p**). However, when CatNMs activity was inhibited by heating the nanomotor suspension at 80°C for 1 h (Fig. S3), the activity-inhibited CatNMs remained homogeneously distributed over time under the same $H_2O_2$ gradient, indicating the absence of directed motion (Fig. 3**n**, S4). The horizontal fluorescence intensity projection shows that particle distribution remains unchanged along the y-axis, further confirming the role of transversal fuel chemoattraction (Fig. S5). Additionally, fluorescence centroid displacement analysis (Fig. 3**q**) reveals that in an $H_2O_2$ gradient with 25 mM concentration difference CatNMs undergo approximately 60 μm displacement along the x-axis (toward the fuel source) over 30 min, whereas the displacement is less than 30 μm under no gradient conditions.

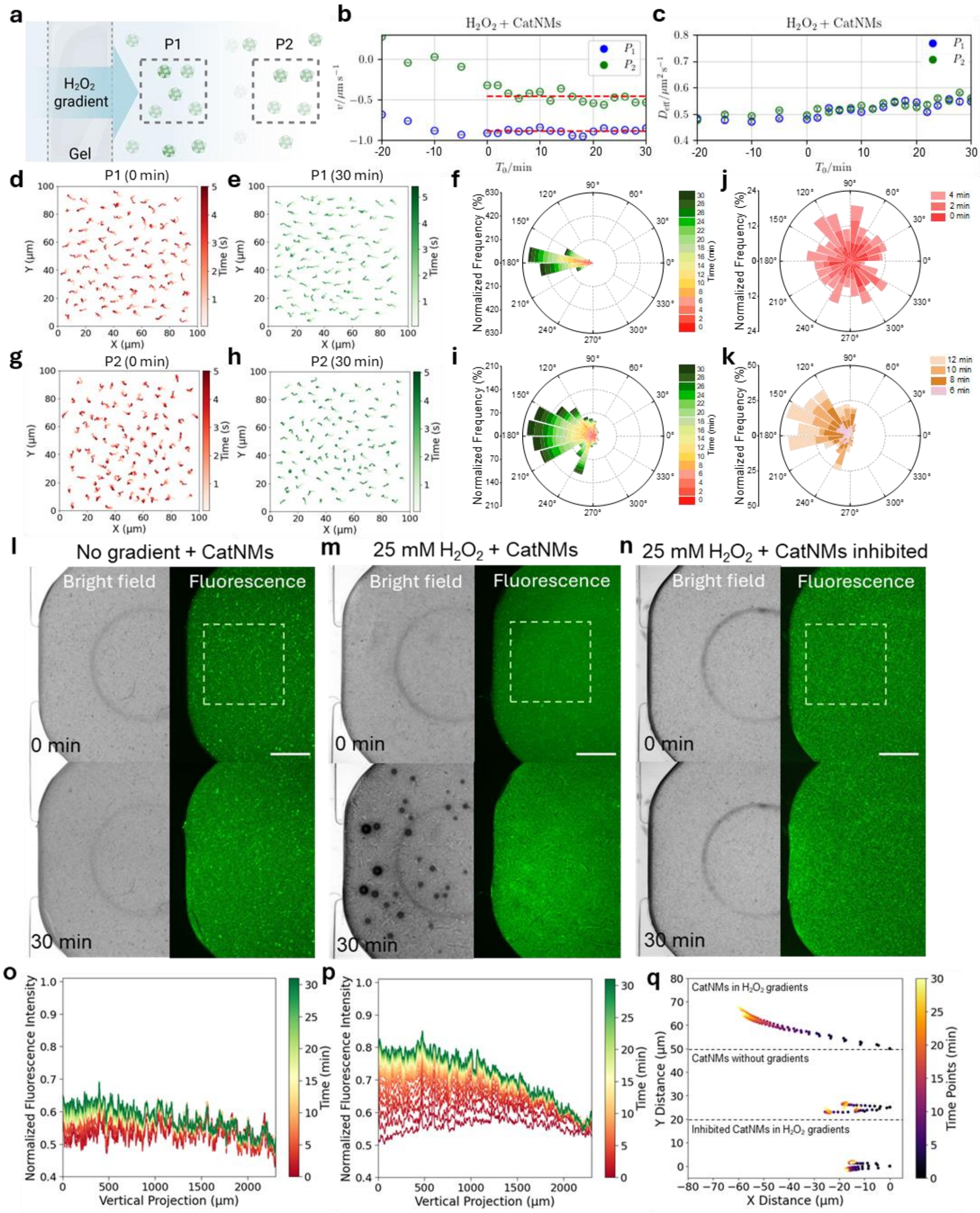

**Fig. 3 Single-particle tracking and collective behavior of CatNMs in $H_2O_2$ gradients. a** Schematic of CatNMs under a $H_2O_2$ gradient. Dashed squares $P_1$ and $P_2$ are the measurement positions for single-particle motion analysis. **b** The velocities of particles at $P_1$ (blue) and $P_2$ (green) are shown from the moment fuel is introduced. Over a period of 20 minutes, a clear relaxation of the velocities toward a plateau is visible (from -20 to 0 minute). The red dashed lines represent the average velocity from 0 to 30 minutes. Data are from N = 10 independent experiments, with approximately 100 particles analyzed per experiment. **c** Effective diffusion coefficients are roughly equal in the two measurement positions. The obtained values are roughly equal in the two measurement positions and remain roughly constant during the relaxation time. After 20 min, a slight increase over time is visible. **d** Representative 5-second trajectories of particles at 0 min and **e** 30 min at $P_1$. **f** Polar map of nanomotor orientations within 5 s at every 2-minute intervals at $P_1$. Approximately 100 particles were analyzed. **g**

Representative 5-second trajectories of particles at 0 min and **h** 30 min at $P_2$. **i** Polar map of nanomotor orientations within 5 s at every 2-minute intervals at $P_2$. Approximately 100 particles were analyzed. **j** The polar map of particle orientations within 0-4 min, and **k** within 6-12 min. **l-n** Representative bright-field and fluorescence images of CatNMs filled in the right channel of the chip under different conditions: **l** no gradient (0.1× PBS solution), **m** $H_2O_2$ gradient (25 mM), and **n** activity-inhibited CatNMs in $H_2O_2$ gradient (25 mM), captured at 0 and 30 minutes. Scale bars: 1 mm. **o**, **p** Normalized fluorescence intensity profiles derived from the squares in **l**, **m**, respectively. **q** Time-dependent displacement of the fluorescence centroid of CatNMs across different conditions. N = 3 independent experiments.

**Communication signals released by GOxNMs.** Chemical signals play the key role in the coordination of the two populations of nanomotors. The experimental configuration for studying chemical communication between the two nanomotors is illustrated in Figs. 4**a**, **b**. GOxNMs were embedded in the agarose gel and introduced into the middle channel of the microfluidic chip, visualized through the intrinsic green fluorescence of FITC (Fig. 4**a**). Upon introduction of glucose solution into the left channel, diffusion established a glucose gradient across the central channel, where GOxNMs catalyzed the oxidation of glucose to gluconolactone and $H_2O_2$, thereby generating a self-sustaining $H_2O_2$ gradient capable of recruiting CatNMs (Fig. 4**b**).

To enable real-time $H_2O_2$ monitoring, we used a colorimetric assay based on the oxidation of 2,2'-azino-bis(3-ethylbenzothiazoline-6-sulfonic acid) (ABTS) by $H_2O_2$ in the presence of horseradish peroxidase (HPR)[27]. As shown in Figure 4**c** and video S3, a detection solution containing ABTS and HRP was added into the right channel. As glucose diffuses through the middle channel, the enzymatically generated $H_2O_2$ diffuses into the right channel, where it triggers HRP-catalyzed ABTS oxidation, producing the ABTS radical cation ($ABTS^{+\cdot}$) with a characteristic blue-green color. Thus, the color gradient observed in the right channel reflects the established secondary $H_2O_2$ gradient. Based on this coloring reaction, the apparent Michaelis constant (${}^{app}K_m$) and maximum rate ($V_{max}$) for GOxNMs were determined to be 6.10 ± 3.30 mM and 0.15 ± 0.02 µmol $min^{-1}$ $mg^{-1}$, respectively, from Michaelis-Menten kinetic analysis (Fig. 4**d**, S6).

We further quantified the produced $H_2O_2$ concentration. In the control group (Figs. 4**e**, **f**), higher $H_2O_2$ concentrations result in more colored $ABTS^{+\cdot}$, independent of glucose levels. However, when GOxNMs was introduced into different groups containing varying glucose concentrations, distinct color change was observed (Fig. 4**g**), reflecting enzyme-mediated $H_2O_2$ generation. In 30 min, all samples transitioned to dark blue, likely due to reaction saturation (Fig. 4**h**). $H_2O_2$ production increased linearly with both GOxNM and glucose concentration (Fig. 4**i-l**). In 5 mM glucose, 0.5 mg/mL GOxNMs yield ~0.52 mM $H_2O_2$ within 5 min, whereas 2 mg/mL GOxNMs generated ~0.86 mM $H_2O_2$ under the same conditions. At 1 mg/mL GOxNMs, $H_2O_2$ production reached 2.0 mM in 25 mM glucose and 2.4 mM in 100 mM glucose. However, at 200 mM glucose, $H_2O_2$ production plateaued at 1.73 mM, likely due to enzyme inhibition at high substrate concentrations. This $H_2O_2$ production level matches the amount needed for enhanced CatNMs diffusion. Single-particle motion analysis of CatNMs in 0–5 mM $H_2O_2$ (Fig. S7) shows apparent enhanced diffusion at 5 mM $H_2O_2$, a concentration comparable to millimolar levels generated by GOxNMs in 100 mM glucose, while staying strictly below the single colloid Einstein–Stokes relation of around 1.01 $\mu m^2 s^{-1}$, presumably due to cluster formation kinetics between colloids[19]. These results support the feasibility of GOxNMs–CatNMs communication under glucose gradients.

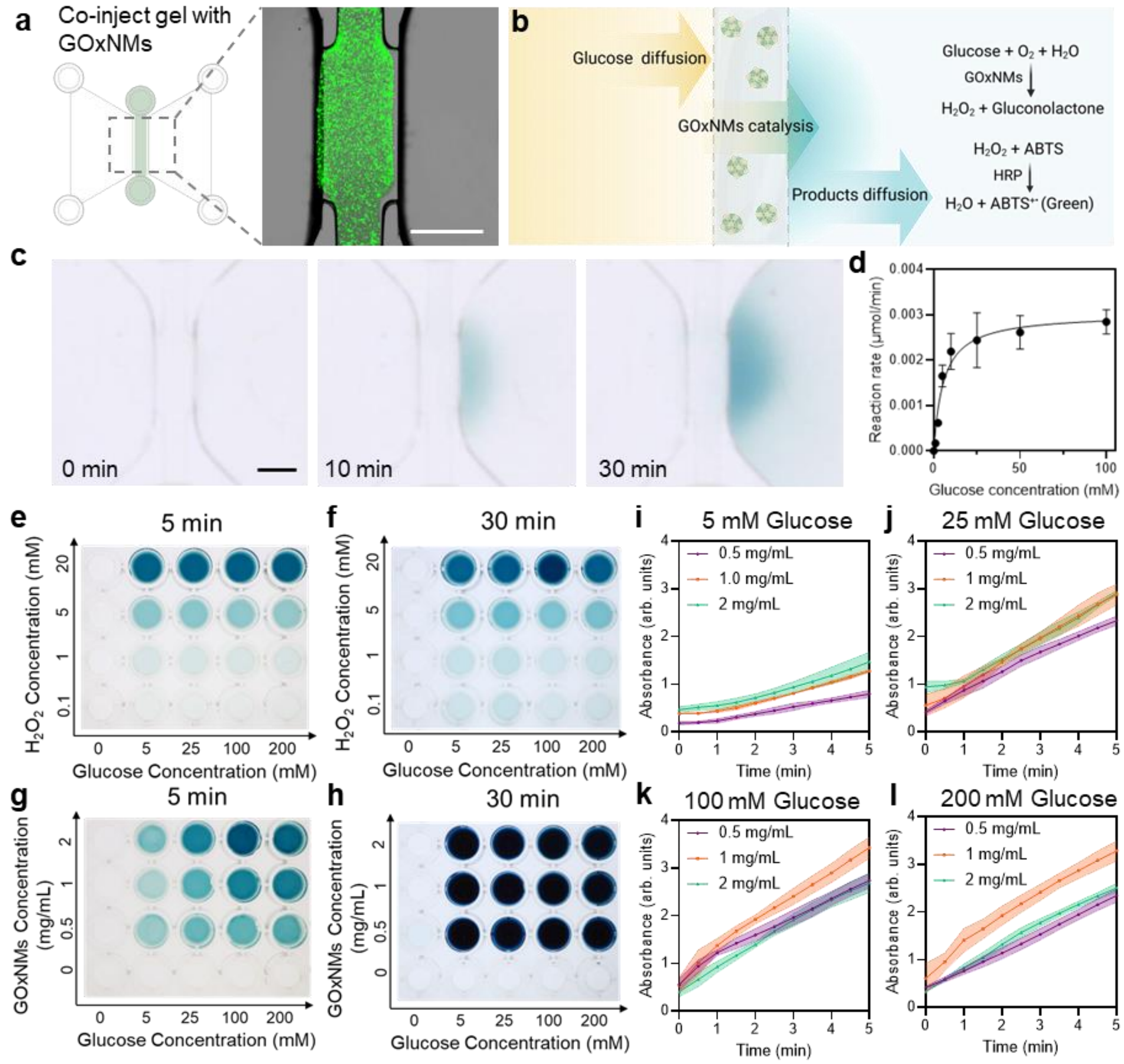

**Fig. 4 Communication signals released by GOxNMs.** **a** Schematic and merged fluorescence/gray image of GOxNMs embedded within gel in the middle channel. Scale bar: 1 mm. **b** Schematic of the communication signals released by glucose-GOxNMs catalysis. The produced secondary $H_2O_2$ gradient is visualized by ABTS/HRP colorimetric reactions. **c** Temporal evolution of the $H_2O_2$ gradient generated from diffusing glucose through embedded GOxNMs. **d** Michaelis–Menten kinetics of GOxNMs in varied concentrations of glucose. N=3 independent experiments. **e-h** ABTS/HRP-based colorimetric detection of $H_2O_2$. The color changes only with $H_2O_2$ concentration. Under the same $H_2O_2$ concentration, the color reaction does not change with varying glucose concentrations or over time, as shown in **e** after 5 min and **f** after 30 min. **g** Under the same concentration of GOxNMs, the color changes with glucose concentration due to different amount of $H_2O_2$ produced by chemical reactions between the nanomotors and varying glucose concentrations. **h** After 30 min, the produced $H_2O_2$ saturates the color reaction, resulting in the same dark green color across all wells. **i-l** Time-dependent absorbance spectra of oxidized ABTS, indicating $H_2O_2$ production kinetics by GOxNMs (0.5-

2 mg/mL) at glucose concentrations of **i** 5 mM, **j** 25 mM, **k** 100 mM, and **l** 200 mM. N=3 independent experiments.

**Chemical communication between CatNMs and GOxNMs.** After establishing the secondary $H_2O_2$ gradient generated by glucose and GOxNMs, we analyzed the motion of CatNMs under this communication signal (Fig. 5**a**). As in the glucose-GOxNMs and $H_2O_2$-CatNMs systems, the CatNMs velocity towards the secondary $H_2O_2$ fuel gradient reaches a plateau value at around 20 min (Figs. 5**b**), and the obtained effective diffusion constants for $P_1$ and $P_2$ display a consistent upward trend and remain close in magnitude (Fig. 5**c**). However, the directionalities of particle trajectories reveal that CatNMs exhibit distinct motion at $P_1$ and $P_2$. In $P_1$, particle trajectories (Figs. 5**d**, **e**) and their directionality (Fig. 5**f**) demonstrate overall directed motion toward the fuel side (video S4). In $P_2$, particle trajectories and their directionality are time-dependent (video S5). These trajectories show no visible directed motion toward either side throughout the recording time (Figs. 5**g**, **h**). Time-evolution polar maps (Fig. 5**i**) indicate that during the first 6 minutes, CatNMs show equal preference for the left and right sides, after which they gradually exhibit biased motion toward the right side, away from the fuel source.

Communication between the two nanomotor populations proved effective, as CatNMs show collective migration toward the self-generated $H_2O_2$ gradients (Figs. 5**j**-**l**, video S6). Under a glucose gradient with 25 mM concentration difference, fluorescence images reveal no notable nanomotor aggregation (Fig. 5**j**), and the fluorescence intensity distribution remains nearly uniform over time (Fig. 5**m**). In contrast, 100 mM glucose concentration difference induces pronounced CatNMs aggregation near the gel side, as evidenced by both fluorescence images (Fig. 5**k**) and the corresponding normalized intensity profiles (Fig. 5**n**). This heterogeneous distribution confirms chemoattraction towards regions of higher $H_2O_2$ concentration. However, at 200 mM glucose (Fig. 5**l**), chemoattraction toward the fuel source persists, while a distinct repulsion zone emerges near the gel surface. This produced an arched fluorescence distribution profile (Fig. 5**o**). Although both 100 mM and 200 mM glucose conditions drive collective motion toward the fuel source, the 100 mM group demonstrates stronger aggregation. Fluorescence centroid displacement analysis across three experimental replicates confirms these trends (Fig. S8). To clarify this counter-intuitive nanomotor distribution, we zoomed in on the repulsion zone and analyzed the motion behavior of CatNMs in this area (Figs. S9, S10). Both the velocity profile and nanomotor orientation map demonstrate that CatNMs can be repelled in the near gel zone. The repulsion of CatNMs from the source region which contains GOxNMs can be explained via the non-reciprocal interaction between the GOxNMs inside the gel and the CatNMs inside the chamber. Using the values and the signs of the mobilities, as well as reactions rates and concentrations, we conclude that our system falls well within the region of the phase diagram of Ref.[11] that predicts active phase separation and segregation (see Extended Fig. E1 and Fig. S11 for further discussions regarding experimental probe of the depletion zone).

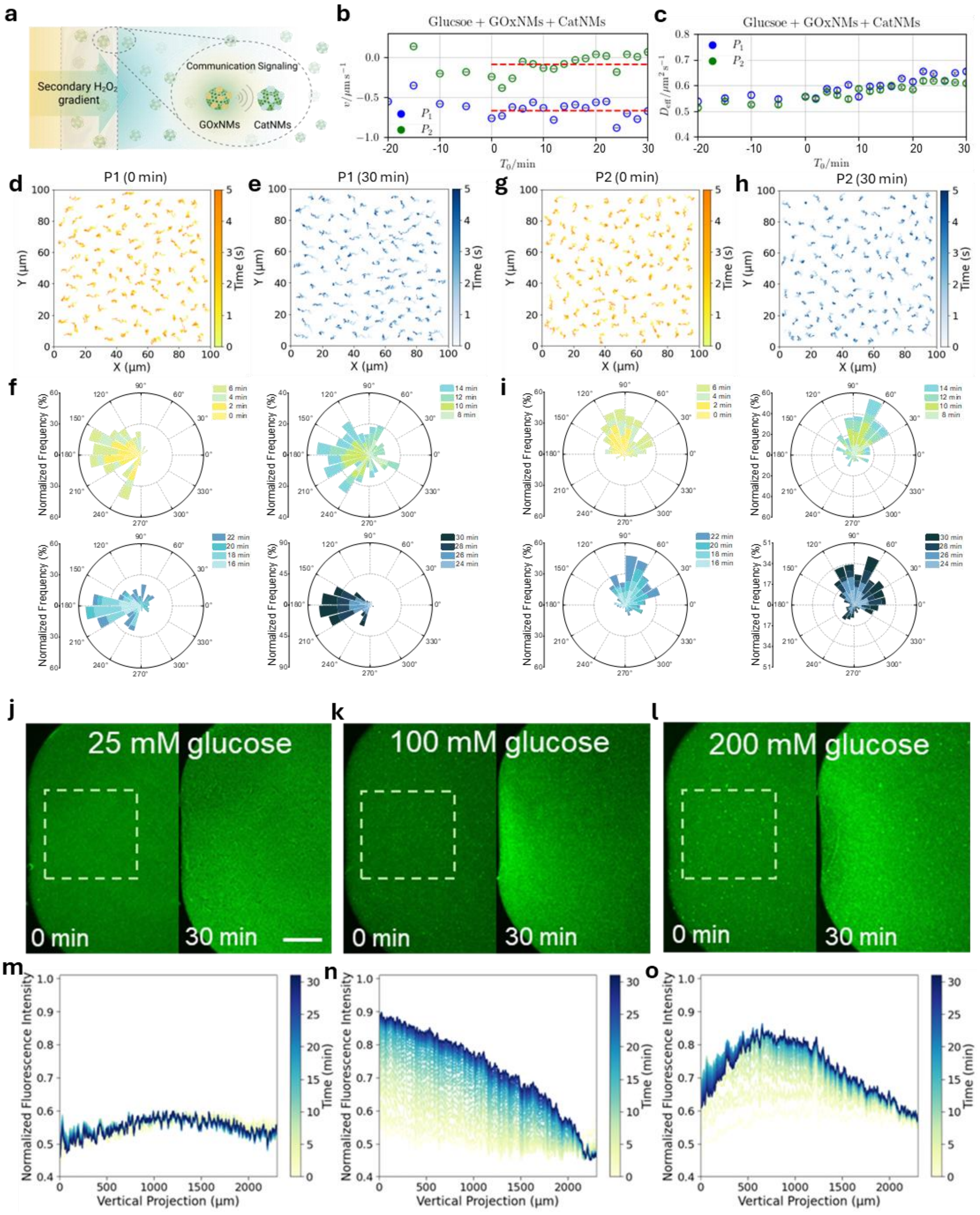

**Fig. 5 Single-particle tracking and collective motion of CatNMs in $H_2O_2$ gradients. a** Schematic of CatNMs under a secondary $H_2O_2$ gradient produced by the embedded GOxNMs, a communication signaling process between the two nanomotors. **b** The velocities of particles at $P_1$ (blue) and $P_2$ (green) are shown from the moment fuel is introduced. Over a period of 20 minutes, a clear relaxation of the velocities toward a plateau is visible (from -20 to 0 minute). The red dashed lines represent the average velocity from 0 to 30 minutes. Data are from N = 10 independent experiments, with approximately 100 particles analyzed per experiment. **c** Effective diffusion coefficients are roughly equal in the two measurement positions. The obtained values are roughly equal in the two measurement positions and remain roughly constant during the relaxation time. After 20 min, a slight increase over time is visible. **d** Representative 5-second trajectories of particles at 0 min and **e** 30 min at $P_1$. **f** Polar map of nanomotor orientations within 5 s at every 2-minute intervals at $P_1$. Approximately 100 particles were analyzed. **g** Representative 5-second trajectories of particles at 0 min and **h** 30 min at $P_2$. **i** Polar map of nanomotor orientations within 5 s at every 2-minute intervals at $P_2$. Approximately 100 particles were analyzed. **j-l** Representative fluorescence images of CatNMs filled in the right channel of the chip under different conditions: **j** Glucose gradient (25 mM), **k** Glucose gradient (100 mM), and **l** Glucose gradient (200 mM), captured at 0 and 30 minutes. Scale bar: 1 mm. **m-o** Normalized fluorescence intensity profiles derived from the squares in **j-l**, respectively.

**Control experiments in support of our proposed mechanism**

**1. Convective flow.** Our recent studies revealed that convective flow driven by concentration gradients may lead to directed particle motion[14]. However, this mechanism does not appear to be applicable to our current system. If it were, the elevated glucose concentration (Fig. 2**j-l**) would be expected to reduce nanomotor aggregation, as the higher mass density on the left side should initiate a counterclockwise fluid flow that repels nanomotors from aggregating at the bottom plane. We further verified our speculation by introducing high concentration of urea solution (300 mM), which is also a non-electrolyte solution so as not to introduce any other ion effect. As shown in Fig. S12, the CatNMs fluorescence centroid displacement remains in a range similar to the control groups in Fig. 3**l**. These results confirm that the agarose gel embedded in the confined middle channel serves as a gate for slowly releasing fuel, effectively inhibiting a transversal convective flow.

**2. Passive phoretic transport.** To establish the order of magnitude and sign of the passive (i.e., non-collective and non-active) phoretic mobilities, we have studied the phoretic movement of individual NMs in the absence of their own substrates and products. As shown in Extended Fig. E2, we have probed the motion of CatNMs caused by glucose gradients with 100 mM of concentration difference. We observe that CatNMs move away from the glucose source reflecting a positive phoretic mobility, which we infer to be $1.8\times10^{-12}$ $m^5$ $mol^{-1}$ $s^{-1}$. This is a reasonable value, which consistent with a Derjaguin length of ~1 nm, as expected[28]. Importantly, this single-particle phoretic mobility is two orders of magnitude smaller than the effective collective mobilities we infer from the chemically active cases, verifying that many-body effects are important for the overall observed behavior reported in this work.

**3. Gel-particle interactions.** Note that due to the porous structure of the agarose gel, some NMs may be trapped at the gel interfaces. However, this phenomenon is not attributed to electrostatic interactions between nanomotors and the agarose gel. The agarose gel contains a low concentration (0.12%) of negatively charged sulfate groups, and the enzyme-functionalized nanomotors also exhibit a negative zeta potential[29]. As a result, electrostatic attraction between gel and nanomotors is unlikely to be the cause of attractive aggregation, ruling out a possible mechanism arising from the gel-particle interactions.

**Theoretical framework for the study of the collective behavior of ENMs used for inference.** Every ENM acts as an active sink of its substrate and an active source of its product, thereby creating and sustaining gradients of substrates and products as they move while responding to gradient due to other ENMs (be it of the same species or the other species) as well as externally generated gradients of the substrate in the microfluidic system. The phoretic velocities can be understood, at a coarse-grained level, as the response of the ENM via a phoretic mobility $\mu$ to the local concentration gradients (Fig. 6**a**). There will be different phoretic mobilities for the response to gradients of substrates and products for each ENM. A stationary velocity profile implies stationary fuel and product gradients within the right chamber of the chip. Given that the plateau values are nonzero, fuel consumption, product generation, and fuel influx must balance to create a nontrivial concentration profile. This interpretation is supported by the differing plateau velocities in the near and far regions, which are incompatible with a constant gradient. In particular, while the velocities at $P_1$ and $P_2$ are similar in magnitude for $H_2O_2$ + CatNMs (Fig. 3**b**), they differ substantially for the glucose + GOxNMs and the glucose + GOxNMs + CatNMs setups (Figs. 2**b**, 5**b**).

The experimentally observed velocity profiles, sampled at $P_1$ and $P_2$, fall into two different classes (Fig. 6**b**). Because of no flux boundaries at the chip wall, the phoretic velocities vanish, implying that for all setups a linear velocity profile is not possible when extrapolated. Instead, the glucose-based systems show a sublinear profile, resulting in a decreased velocity far from the gel, whereas the $H_2O_2$ system shows superlinear profile, for which the velocities remain similar in the two regions. To understand these two categories of behavior mechanistically, we used a nonlinear framework that includes the phoretic response of the ENMs to the substrates and products as well as the nonlinear dynamics of the chemicals themselves as coupled back to the ENMs. More precisely, the model includes the degradation of the substrate $c_1$, the production of the product $c_2$, and the drift of the nanomotor density $\rho$ due to phoretic interactions mediated by the nondimensional phoretic mobilities $\mu_{1/2}$. The full model and the nondimensionalization are described in the Methods section. The screening length associated with the overall degradation of the chemicals is estimated to be of the order of millimeters, comparable to the system size $L$=6 mm. For different combinations of substrate and product mobilities, the model reproduces both the super- and sublinear velocity profiles (Fig. 6**b**), which represent signatures of the type of collective interaction the ENMs have with their substrates and products, as we discuss below.

A broad parameter scan over the four different combinatorial possibilities of the signs of the two phoretic mobilities (namely, ++,+-,-+,--) reveals that the phase diagram is dominated by the sublinear regime (blue region in Fig. 6**c**), while the superlinear regime (red regions in Fig. 6**c**) can only be found in the $\text{sign}(\mu_1) = \text{sign}(\mu_2)$, $\mu_1<\mu_2$ sectors, i.e. equal signs for substrate and product phoresis with product phoresis being dominant. Note that without product phoresis, i.e. $\mu_2 = 0$, no superlinear velocity profiles are possible. This is an important new feature that highlights the decisive role of phoretic response to products of an enzymatic reaction on the outcome of their collective self-organization. This is to be contrasted with chemotaxis towards the substrates, which is a necessary ingredient in the self-organization but is not sufficient to engineer complex tunable behavior in the system. The white diagonal corresponds to the case of equal phoretic mobilities, resulting in a linear velocity profile. Near $\mu_1 = 0$ a band of dark red and blue is visible, corresponding to the linear interpolation having close to zero slope, resulting in large positive or negative roots, depending on the relative magnitude of the two velocities. The inversion symmetry of the phase diagrams is a consequence of the velocity profiles changing signs as we go from ($\mu_1$, $\mu_2$) to ($-\mu_1$, $-\mu_2$). Depending on the relative magnitude and signs of the phoretic mobilities, the boundary layer at the gel is either desaturated (red and orange lines in Fig. 6**d**) or saturated (magenta and green lines in Fig. 6**d**). In addition, in the superlinear regime, a local depletion/aggregation of particles close to the boundary layer is visible. Velocity profiles (Fig. 6**e**) were obtained at the color-coded points in the phase diagram Fig. 6**a**. A clear distinction between the super- and sublinear regimes is visible.

Upon increasing the ENM density, the system enters a more strongly screened regime (Fig. 6**f**), introducing regions in which the velocity profiles deviate strongly from the linear case (dark blue and red regions). In this regime, the nonmonotonicity of the density profiles is exaggerated (Fig. 6**g**). The velocity profiles also become more exaggerated (Fig. 6**h**), with the sublinear profiles crossing $v = 0$.

While the nanomotor density reaches a stationary state (Fig. 6**i**), the fuel and product concentrations only reach stationary-gradient states (Figs. 6**j**, **k**). This occurs because the continuous influx through the gel increases the global concentration, whereas the boundary conditions, together with the nanomotor aggregation near the gel, fix the gradient. Consequently, the ENM velocities in the two measurement positions reach plateau values (Fig. S13), which is consistent with the experimental observations and also generically observed in a wide range of parameter combinations. This highlights the importance of nontrivial collective dynamics with regards to the designability of the overall behavior of such ENM chemotactic assays.

In the stationary-gradient states, the chemical concentrations split into a time-independent spatial part and a spatially independent global shift. This global increase in concentration can be linked to the relative enhancements of the diffusion coefficient after 20 min, visible in Figs. 2, 3, and 5 (panel **c**). As the concentration increases, the particles become more active, resulting in the breaking of clusters into smaller entities.

We note that the nanomotor density profile (Fig. 6**i**) in this simple implementation of the model does not reproduce the same curvature as the experimentally measured fluorescence profiles (Figs. 2**m-o**, 3**o-q**, 5**m-o**). We expect a model that simultaneously incorporates both ENM species to account for the curvature change of the full profile.

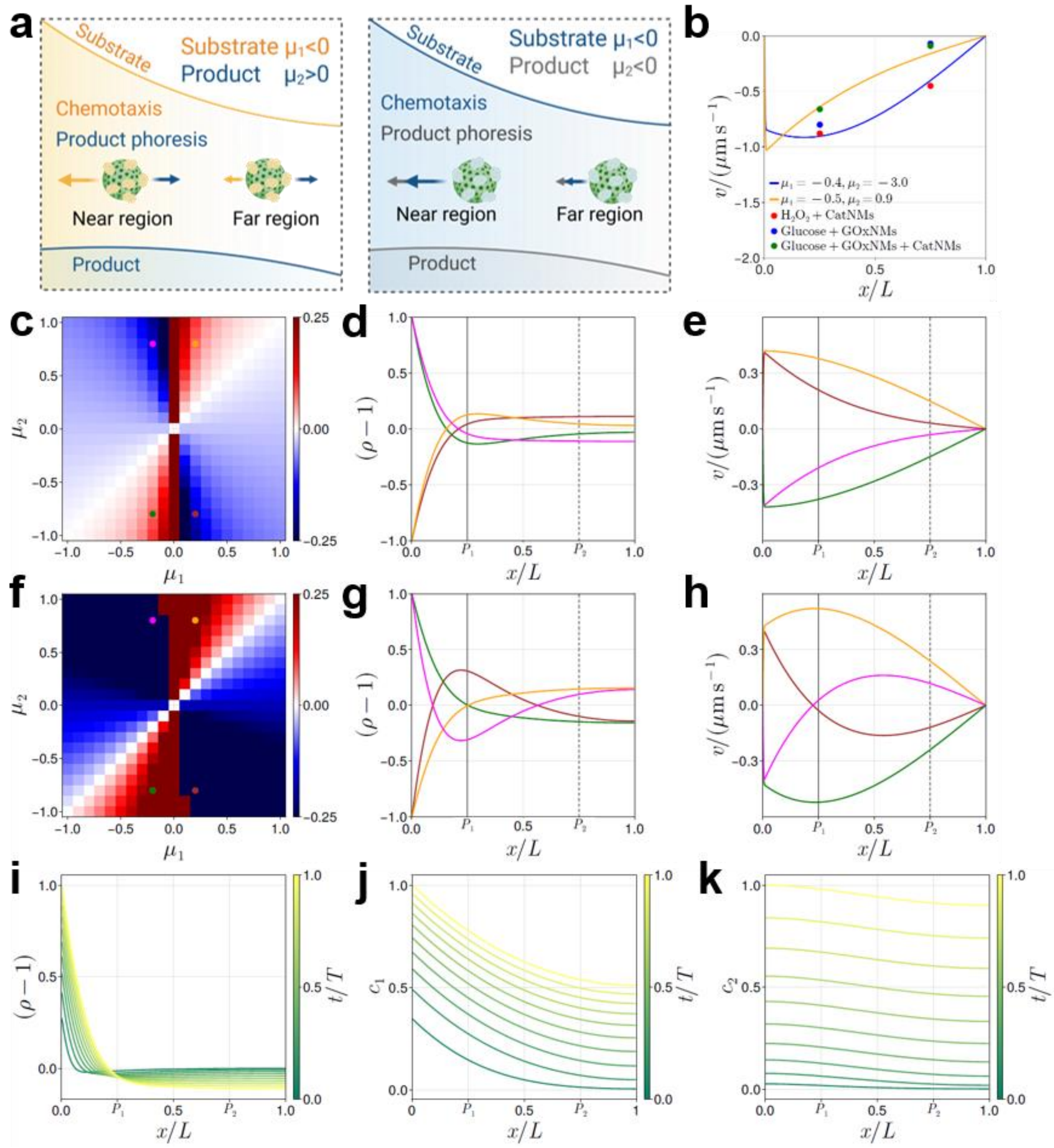

**Fig. 6 Numerical analysis of nanomotors in fuel gradients. a** Schematic of GOxNMs in a glucose gradient and CatNMs in a $H_2O_2$ gradient. The phoretic mobilities of the substrate ($\mu_1$) and the product ($\mu_2$) influence the collective motion of nanomotors. The signs of the mobilities will be controlled by the interaction of the surface and the solute particles, i.e. if the interaction potential is repulsive or attractive. When $\mu_1 < 0$ and $\mu_2 > 0$, nanomotors in the near region are primarily attracted, while the motion of those in the far region results from a trade-off between substrate and product phoretic response. When both $\mu_1 < 0$ and $\mu_2 < 0$, nanomotors in both the near and far regions are attracted. **b** Experimentally obtained velocities (colored dots) and numerically obtained velocity profiles (colored lines), sampled at the two measurement positions $P_1$ and $P_2$. Due to the boundary condition of vanishing velocity at the chip wall, the three setups fall into two classes. The sublinear class (Glucose + GOxNMs, Glucose + GOxNMs + CatNMs) and the superlinear class ($H_2O_2$ + CatNMs). **c** Numerical parameter scan for different combinations of substrate $\mu_1$ and product $\mu_2$ phoretic mobilities. The color shows the root of the linear interpolation between the velocities at the two measurement positions $P_1$ and $P_2$, shifted by the domain size. The blue regions correspond to the sublinear class, while the red regions correspond to the superlinear class. **d** Steady state density profiles at different mobility combinations given by the same color dots in **c**. **e** Steady state velocity profiles at the same mobility combinations as in **d**. The inversion symmetry of **c** corresponds to a mirror symmetry of **e** around v=0. **f-h** Same as **c-e**, but for an increased nanomotor density, corresponding to a strongly screened regime. The dark red and dark blue regions around $\mu_1 = 0$ have grown in size. The dark blue regions correspond to configurations in which the near-fuel velocity is lower than the far-fuel velocity. **i** Normalized and shifted particle density (ρ) profiles, for the $\mu_1 = -0.3$, $\mu_2 = -0.8$ (green dot in **c**) phoretic mobility combination. Different times are color-coded with time increasing from green to yellow. **j-k** Normalized substrate and product concentrations over time.

## Conclusions

In conclusion, this study establishes a system that exhibits chemical communication between two types of ENMs based on cascade reactions. We have developed a carefully designed experimental set up that allows us to infer spatiotemporally resolved patterns of behavior under different gradient conditions. Based on the motion analysis at single-particle and collective level, we have probed the chemotactic behavior of GOxNMs and CatNMs in response to their substrates and products. Building on this, we developed a two-step chemoattraction strategy: a glucose gradient first attracts GOxNMs, which generate a localized $H_2O_2$ gradient that further attracts CatNMs. The concrete experimental results exhibit a trend that for glucose + GOxNMs (+ CatNMs) and $H_2O_2$ + CatNM setups, the velocity profiles can be sublinear or superlinear, from which we have uncovered the key role of product phoresis in determining the combined chemotactic behavior of ENMs that are connected to each other in a cascade chemical reaction network. This work represents a significant step toward creating synthetic systems capable of chemical communication and coordinated behavior, core features of collective biological processes.

## Methods

### Synthesis of FITC-MSNPs

FITC-labeled mesoporous silica nanoparticles (MSNPs) were synthesized according to our previous report with slight modifications[30,31]. First, FITC (Sigma-Aldrich 3326-32-7) was conjugated with 3-aminopropyltriethoxysilane (APTES; Sigma-Aldrich 440140) by adding FITC (2 mg) into 5 mL EtOH and 400 µL APTES. The mixture was stirred for 30 min in dark.

In a three-neck round bottom flask, 35 g triethanolamine (TEOA; Sigma-Aldrich 90279), 570 mg hexadecyltrimethylammonium bromide (CTAB; Sigma-Aldrich 52365), and 20 mL DI water were mixed and stirred at 95°C under reflux for 30 min for homogenization. 250 µL FITC-APTES conjugate and 1.25 mL tetraethyl orthosilicate (TEOS; Sigma-Aldrich 131903) were added dropwise separately through one of the three necks of the round-bottom flask. The reaction was under stirring for 2 h in dark. Then stop heating and wait the temperature to cool down until 60°C. The products were collected by centrifugation (2500 rcf, 5 minutes) and washed three times with ethanol.

The as-prepared products were post-processed for removal of CTAB by resuspending them in a mixture of 30 mL ethanol and 1.8 mL HCl for at least 15 h under stirring and reflux in dark. The FITC-MSNPs were collected by centrifugation (2500 rcf, 5 minutes) and washed three times with ethanol.

**Surface functionalization with enzymes**

The surface of FITC-MSNPs was then modified for further functionalization. 20 mg FITC-MSNPs were resuspended in ethanol and 100 μL APTES were mixed. The mixture was placed in an end-to-end shaker at room temperature for 24 h. The resulting amino-functionalized FITC-MSNPs were collected by centrifugation (5000 rpm, 5 min) and washed four times with ethanol to remove residual APTES. The collected FITC-MSNPs-$NH_2$ were dried for further use.

The prepared FITC-MSNPs-$NH_2$ nanoparticles (1 mg) were resuspended in 1 mL 1× PBS and activated with 50 μL glutaraldehyde (Sigma-Aldrich G6257, 25% in $H_2O$) in an end-to-end shaker for 2.5 hours at room temperature. The activated particles were then collected by centrifugation (5000 rpm, 5 min) and washed four times with 1× PBS. The final pellet was resuspended in 1 mL of 1× PBS. Then 0.8 mg Catalase (Sigma-Aldrich 9001-05-2) or glucose oxidase (Sigma-Aldrich 9001-37-0) was added to the activated FITC-MSNPs-$NH_2$ nanoparticles. The reaction was kept at room temperature in an end-to-end shaker overnight. The resulting catalase-nanomotors (CatNMs) or glucose oxidase-nanomotors (GOxNMs) were collected by centrifugation (5000 rpm, 5 min) and washed three times with 1× PBS. The supernatant was collected for further quantification of the enzyme conjugation. Finally, resuspend the collected enzyme-linked nanomotors in 0.1× PBS (1 mL) and store them in the fridge at 4℃ for future use.

**Preparation of gel-filled microfluidic chip**

1% agarose gel was prepared by dissolving 100 mg of agarose powder (Thermo Fisher, 17850) in 10 mL of 0.1× PBS. The mixture was heated to 100℃ with stirring until the agarose was fully dissolved. Subsequently, the dissolved agarose solution (10 μL) was filled into the middle channel of a commercial microfluidic chip (μ-Slide Chemotaxis; ibidi), with dimensions of 1 mm × 2 mm × 70 μm (length × width × height). The solution was allowed to gel for 2 minutes at room temperature. Finally, the two adjacent channels were filled with 0.1× PBS to maintain the gel network for further use.

**Preparation of enzyme-embedded microfluidic chip**

2% agarose gel was prepared by dissolving 200 mg agarose powder in 10 mL 0.1× PBS. The mixture was heated to 100℃ with stirring until the agarose was fully dissolved. The solution was then cooled to 70℃ and immediately mixed with an equal volume (10 μL) of either a suspension of GOxNMs (10 mg/mL in 0.1× PBS) or enzyme GOx (4 mg/mL in 0.1× PBS). The resulting mixture was filled into the middle channel of the microfluidic chip. Wait until the solution cools down and forms gel network in the channel for 2 minutes. Finally, the two adjacent channels were filled with 0.1× PBS to maintain the gel network for further use.

**Establishment of fuel gradient**

Nanomotor suspensions were first injected into the right channel, while the left channel was filled with 0.1× PBS solutions. After a 20-minute incubation, the left (source) channel was filled with fuel solutions: 0.1× PBS (no-gradient control), glucose (25, 100, or 200 mM in DI water), or 25 mM $H_2O_2$ (in DI water). To maintain consistent osmotic pressure, nanomotors were suspended in PBS solutions at adjusted concentrations: 0.1× PBS for the control and 25 mM $H_2O_2$, 0.54× PBS for 100 mM glucose, and 1× PBS for 200 mM glucose.

**Detection of $H_2O_2$ diffusion**

$H_2O_2$ solutions (5 mM or 25 mM) were introduced into the left channel of six three-channel chips, resulting in 18 independent experiments. $H_2O_2$ concentrations in three right channels were measured every 10 min using a commercial Amplex Red hydrogen peroxide assay kit (Thermo Fisher A12222). Briefly, a gradient of $H_2O_2$ concentration (0-10 μM) was prepared for a standard concentration curve. A working solution with 50 μL of 10 mM Amplex Red reagent solution, 100 μL of 10 U/mL horseradish peroxidase (HRP; Sigma-Aldrich 9003-99-0) solution, and 4.85 mL of 1× reaction buffer was prepared. The solutions in three right channels were extracted after 10-min diffusion, the rest were left for diffusion for further detection. Pipet 50 μL of the standard curve samples or experimental samples into individual wells of a 96-well plate, followed by adding 50 μL of the as-prepared working solution to each of the wells. The mixed solutions were incubated at room temperature for 30 min in the dark, and fluorescence was measured using a microplate reader with excitation at 560 nm and emission detection at 590 nm.

**Bicinchoninic acid (BCA) assay**

BCA assay was performed to quantify the amount of glucose oxidase and catalase linked onto the surface of FITC-MSNPs and FITC-MSMPs (Table S1), with bovine serum albumin (BSA) as the protein standard (Thermo Fisher 23227). BSA solutions at concentrations of 0-2000 μg/mL were prepared. Aliquots of 25 μL from each BSA standards and from the supernatants of GOxNMs, CatNMs, or GOxMPs samples collected during surface functionalization were added in individual wells of a 96-well plate. Then, 200 μL of a light-sensitive working reagent (prepared by mixing 50 parts reagent A with 1 part reagent B) was added to each well. The plate was gently shaken for 30 s and incubated at 37°C for 30 min. Absorbance at 562 nm was measured for both standards and samples. Protein concentrations were determined by comparing sample absorbances to the standard curve, accounting for the initial protein amount added and the residual protein remaining in the supernatant.

**Kinetics measurements of GOxNMs**

Kinetic parameters ($K_m$ and $V_{max}$) were determined at room temperature in a 96-well plate. A series of glucose solutions (0, 1, 2.5, 5, 10, 25, 50, 100 mM) was prepared. For each assay, 200 μL of a mixture containing the specified glucose concentration, 1 mM ABTS, and 0.1 ng/mL fresh HRP was added to the plate. Each concentration was tested in triplicate. The reaction was initiated by adding 20 μL of GOxNMs suspension (1 mg/mL). The absorbance at 416 nm was measured every 11 seconds for 2 hours in a multimode microplate reader (BioTek Synergy HTX) with orbital shaking. The change in absorbance was converted to concentration using a molar attenuation coefficient of $3.6 \times 10^4$ $M^{-1}$ $cm^{-1}$.

**Fluorescence image acquisition and analysis**

The distribution of nanomotors under fuel gradients was recorded using a Thunder inverted fluorescence microscope equipped with a 2.5× objective lens. Bright-field and fluorescence (excitation at 470 nm) images were acquired every minute (1 frame per minute, 1 fpm), starting at the introduction of nanomotors (0 min) and continuing 30 minutes. The normalized fluorescence intensity distribution along the vertical projection was analyzed using the Quantify Tool in Leica Application Suite X software. This tool is designed for point spread function analysis, which involves determining mean gray-scale values in images based on ROIs. To measure the mean gray-scale values, the arithmetical mean of the gray-scale values within a selected area ROI is calculated, either line-by-line (horizontal) or column-by-column (vertical). The evaluation results were displayed by selecting the fluorescence channel, sort ROIs, and either vertical or horizontal average projection.

**Optical video recording and motion analysis**

The diffusion of the self-generated $H_2O_2$ gradient was recorded using a digital camera (Thorlabs, CS165CU/M) equipped with a lens (FUJINON, HF35HA-1S). Glucose solution (100 mM, 60 uL) was filled into the left channel of the enzyme-embedded microfluidic chip, and a mixture of ABTS (Sigma-Aldrich A1888; 4 mM, 60 uL) and HRP (Sigma-Aldrich P8250; 0.4 mg) was introduced into the right channel. Videos were recorded at a frame rate of 15 fps for 30 min.

The collective motion of nanomotors in gradients was recorded using a Leica DMi8 microscope equipped with a high-speed, cooled charge-coupled device camera from Hamamatsu and a 40× objective lens. Due to the continuity of fluid flow in a closed space, particle motion is expected to vary across different $z$-planes. We focused on a height ~100 μm above the bottom plane. To avoid initial disturbances when introducing the nanomotor suspensions, we first filled the right channel with nanomotors. After allowing the nanomotors to stabilize for 20 minutes, we introduced the fuel into the left channel. Particle tracking was then performed at two fixed positions for 50 minutes. The position $P_1$ is a 300 × 300 μm square, approximately 1.3 mm away from the right edge of the middle channel. The position $P_2$ is a 300 × 300 μm square, approximately 3 mm away from the position $P_2$. Videos were recorded at 25 fps for 10 seconds at 5-minute intervals in the initial 20 min and at 2-minute intervals in the following 30 min.

Data processing was performed in two main stages: background removal and single-particle tracking. To remove static background noise, each frame of the video was corrected by subtracting the average intensity image, computed from 100 randomly

selected frames:

$$I_{\text{corrected}}(x,y,f) = I_{\text{raw}}(x,y,f) - \frac{1}{100}\sum_{k=1}^{100} I_k(x',y') \tag{1}$$

where $I_{\text{raw}}(x,y,f)$ is the intensity at pixel $(x,y)$ in frame $f$ of the video sequence, $I_k(x',y')$ is the intensity at pixel $(x',y')$ in the $k^{th}$ selected frame, and $I_{\text{corrected}}(x,y,f)$ is the resulting background-free frame. Following preprocessing, particle trajectories were extracted with a custom-made Python code, based on the open-source Trackpy library[32]. From these trajectories, the mean squared displacement (MSD) was calculated as:

$$\text{MSD}(\Delta t)\ \langle \sum_{i=1}^{n} (x_i(t+\Delta t)\text{-}x_i(t))^2 \rangle \tag{2}$$

where $t$ is the time in seconds, and $n$=2 for 2D analysis. The velocity was then extracted from fitting the MSD curve to:

$$\text{MSD}(\Delta t)=v^2\Delta t^2+4D_c\Delta t \tag{3}$$

where $v$ is the velocity, and $D_c$ is the diffusion coefficient. The errors correspond to the 95% confidence interval (CI 95%).

**ENM directionality**

Polar displacement analysis was performed based on the single-particle tracking data. The displacement components in the $x$ and $y$ directions for each particle were computed as:

$$\Delta x=x(t+\Delta t)\text{-}x(t) \tag{4}$$

$$\Delta y=\big(H\text{-}y(t+\Delta t)\big)\text{-}\big(H\text{-}y(t)\big) \tag{5}$$

where $x(t)$ and $x(t+\Delta t)$ represent the position along the x-axis of a particle at the initial and final time points, respectively, and a time interval of 6 s was used to capture complete displacement vectors. $H$ is the video width (in pixels), and the $y$-coordinate was inverted to maintain the conventional Cartesian coordinate system (with y increasing upwards). The displacement magnitude $r$ and angle $\theta$ (with respect to the positive $x$-axis) were then calculated as:

$$r=\sqrt{(\Delta x)^2+(\Delta y)^2} \tag{6}$$

$$\theta=\text{arctan2}(\Delta y,\Delta x) \tag{7}$$

The collection of angles $\theta$ for all particles was divided into 24 equal angular bins spanning 0º to 360º, corresponding to 15º sectors. This angular distribution was visualized as a polar bar plot.

**Trajectory data analysis**

The trajectory data has been analyzed by using a running average

$$R_n(t_i)=\frac{1}{N_t\text{-}i}\sum_{j=0}^{N_t\text{-}i}\left[x\big(t_i+t_j\big)\text{-}x\big(t_j\big)\right]^n. \tag{8}$$

From these running averages, the mean and variance of the displacements were computed as ensemble averages over the multiple realizations of the experiments

$$\mathrm{Mean}(t_i)=\langle R_1(t_i)\rangle,\ \mathrm{Var}(t_i)=\langle R_2(t_i)\rangle\text{-}\langle R_1(t_i)\rangle^2. \tag{9}$$

The $x$ and $y$ components of the trajectories have been analyzed separately to verify purely diffusive behavior in the $y$ direction. To extract the particle velocities and effective diffusivities from these measurements, the mean and variance were fitted using

$$\mathrm{Mean}(t_i)=v_{T_0}t,\ \mathrm{Var}(t_i)=2D_{\mathrm{eff}}t. \tag{10}$$

**Analytical model**

To model the system analytically, the coupled evolution of the substrate $c_1$, product $c_2$ and particle $\rho$ concentrations were implemented via diffusive terms with diffusion constant $D_c$ and $D_\rho$, a degradation term of the substrate due to Michaelis–Menten kinetics at the particle surface, an equal and opposite production term for the product, an artificial source term for the influx of substrate and substrate and product phoretic drift contributions to the evolution of the conserved particle density. The full model reads

$$\partial_t c_1(\mathbf{r},t) = D_c\boldsymbol{\nabla}^2 c_1(\mathbf{r},t) - \frac{k_e N_e c_1(\mathbf{r},t)\rho(\mathbf{r},t)}{K_M + c_1(\mathbf{r},t)} + s(\mathbf{r}), \tag{11}$$

$$\partial_t c_2(\mathbf{r},t) = D_c\boldsymbol{\nabla}^2 + \frac{k_e N_e c_1(\mathbf{r},t)\rho(\mathbf{r},t)}{K_M + c_1(\mathbf{r},t)}, \tag{12}$$

$$\partial_t \rho(\mathbf{r},t) = D_\rho\boldsymbol{\nabla}^2\rho(\mathbf{r},t) + \boldsymbol{\nabla}\cdot\Big(\rho(\mathbf{r},t)\boldsymbol{\nabla}\big(\tilde{\mu}_1 c_1(\mathbf{r},t) + \tilde{\mu}_2 c_2(\mathbf{r},t)\big)\Big), \tag{13}$$

with $V_{\max}=k_e N_e$ giving the consumption rate per enzyme $k_e$ times the number of enzymes at the surface of the particle $N_e$, the Michaelis constant $K_M$ and the phoretic mobilities $\tilde{\mu}_1$ (substrate), $\tilde{\mu}_2$ (product). The nondimensional form of the model in 1D is given by

$$\partial_t c_1(x,t) = \partial_x^2 c_1(x,t) - \frac{c_1(x,t)\rho(x,t)}{1+c_1(x,t)} + \tilde{\gamma}(x), \tag{14}$$

$$\partial_t c_2(x,t) = \partial_x^2 c_2(x,t) + \frac{c_1(x,t)\rho(x,t)}{1+c_1(x,t)}, \tag{15}$$

$$\partial_t \rho(x,t) = \Delta\partial_x^2\rho(x,t) + \partial_x\Big(\rho(x,t)\partial_x\big(\mu_1 c_1(x,t) + \mu_2 c_2(x,t)\big)\Big), \tag{16}$$

with the nondimensional phoretic mobilities $\mu_{1/2}=\tilde{\mu}_{1/2}K_M/D_c$, the nondimensional diffusion constant $\Delta=D_\rho/D_c$, the screening length $\kappa_c^{-1}=\sqrt{K_M D_c/(k_e N_e \bar{\rho})}$, the screening time $\tau=\kappa_c^{-2}/D_c$ and the average particle density $\bar{\rho}=\int_0^L d\,x\rho(x,t)/L$ . The nondimensional source term is defined by its strength $\gamma = \int_{-\epsilon}^{\epsilon} d\, x\tilde{\gamma}(x)$, where $\tilde{\gamma}(x)$ has a strong localization in the interval $[-\epsilon, \epsilon]$. The model is considered on a domain $[0, L]$, where $L$ is the domain size in units of the screening length. At the boundary $x$=$L$, the model is governed by zero flux conditions, representing the hard wall of the chip. At the boundary $x$=$0$, the particle density, as well as the product concentration are governed by zero flux conditions as well, while the substrate flux is set by the source term.

**Numerical solution of the dynamics**

To solve the model numerically, the domain was artificially mirrored at $x$=0, to obtain a $2L$ periodic model, which can then be solved using a pseudospectral method with exponential time stepping[33].

**Analytic steady gradient solution**

By transforming the chemical fields into the total and difference concentrations $\Psi_{\pm}(x,t)=\big(c_1(x,t) \pm c_2(x,t)\big)/2$ , the model allows for an analytic stationary $\rho$ solution in the unscreened regime, i.e. $L \ll 1$, by using a stationary gradient ansatz $\Psi_{\pm}(x,t)$=$\psi_{\pm}(x)$+$T_{\pm}(t)$. The temporal parts $T_{\pm}(t)$ give global shifts with a constant rate. The stationary $\rho$ profile is given by the implicit integral form

$$x(\xi)= \pm \int_{\xi}^{\xi} \frac{d\xi'}{\xi'\sqrt{(b\text{-}a/L)ln\xi'\text{-}b\rho_L(\xi'\text{-}1)}} + L, \quad (17)$$

with $\xi$=$\rho(x)/\rho_L$, $\rho_L$=$\rho(L)$, $a$=$\gamma(\mu_+ + \mu_-)/(2\Delta)$, $b$=$2\mu_-/\Delta$ and $\mu \pm = \mu_1 \pm \mu_2$.

## Acknowledgements

The research leading to these results has received funding from the grants PID2021-128417OB-I00 and PDC2022-133753-I00 funded by MCIN/AEI/ 10.13039/501100011033 and by "EDRF, EU" (Bots4BB project, BiOrganiBOTS). Additional funding was provided by the European Research Council (ERC) under the European Union’s Horizon 2020 (Grant Agreement No. 866348, i-NanoSwarms) and from the “la Caixa” Foundation under Grant Agreement No. LCF/PR/HR21/52410022. S.S. also acknowledges the “Constantes y Vitales” 2023 prize. The IBEC team wishes to thank the CERCA programme of the Generalitat de Catalunya, the Secretaria d'Universitats i Recerca del Departament d'Empresa i Coneixement de la Generalitat de Catalunya through the project 2021 SGR 01606, and the "Centro de Excelencia Severo Ochoa", funded by Agencia Estatal de Investigación (CEX2018-000789-S). S.S. and S.C. acknowledge the Predoctoral AGAUR-FI Joan Oró grant (2023 FI-1 00654) funded by “Secretaria d'Universitats i Recerca del Departament de Recerca i Universitats de la Generalitat de Catalunya” and by European Social Fund Plus. We acknowledge support from the Max Planck School Matter to Life which is jointly funded by the Federal Ministry of Education and Research (BMBF) of Germany and the Max Planck Society.

## Competing Interests

The authors declare no competing interests.

## Data Availability

The authors declare that all the raw data is available from the corresponding author upon reasonable request.

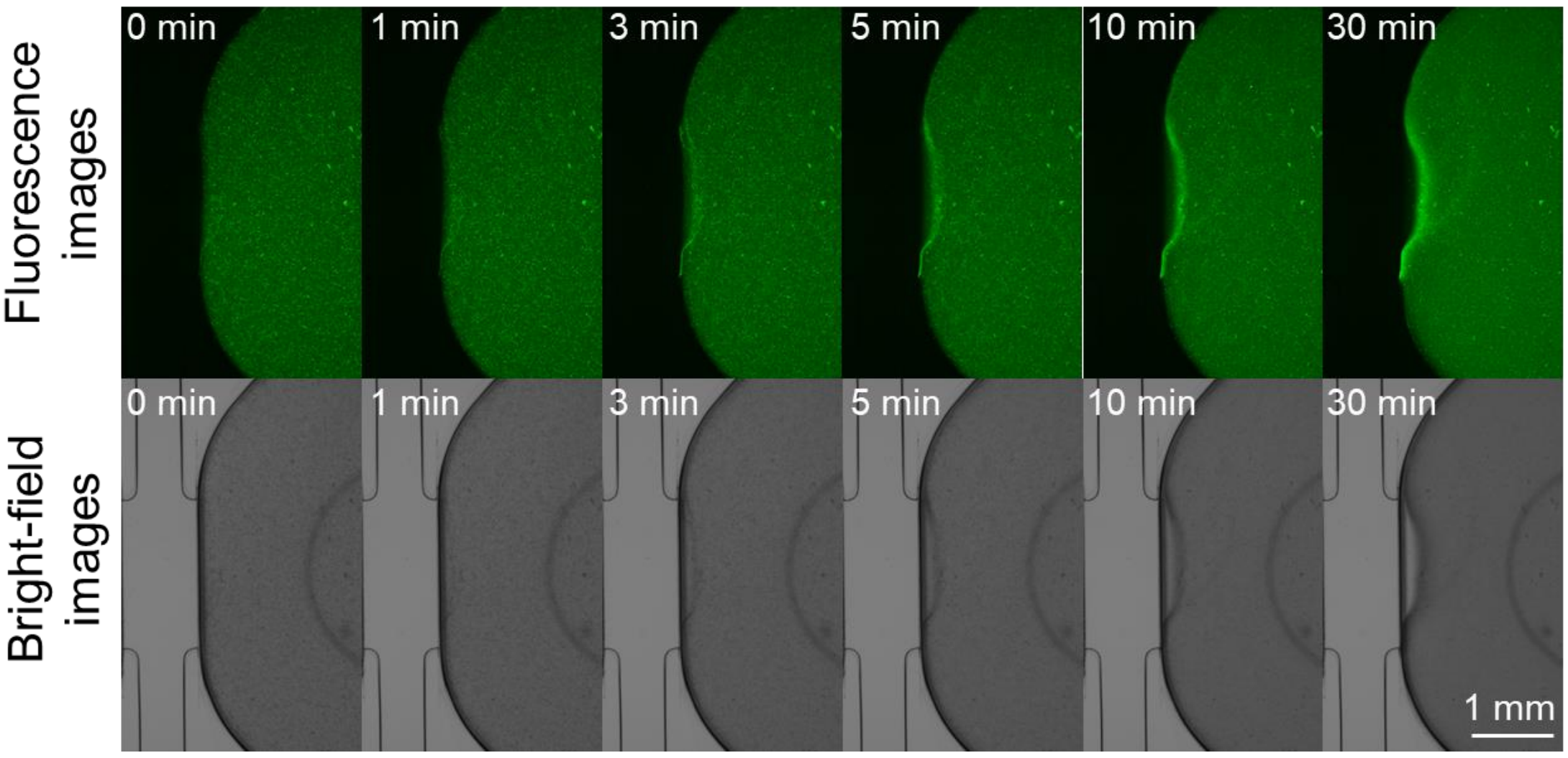

**Extended Fig. E1 Exclusion zone formed near the gel surfaces.** In a system prepared in DI water, nanomotors near the gel side are repelled from the exclusion zone.

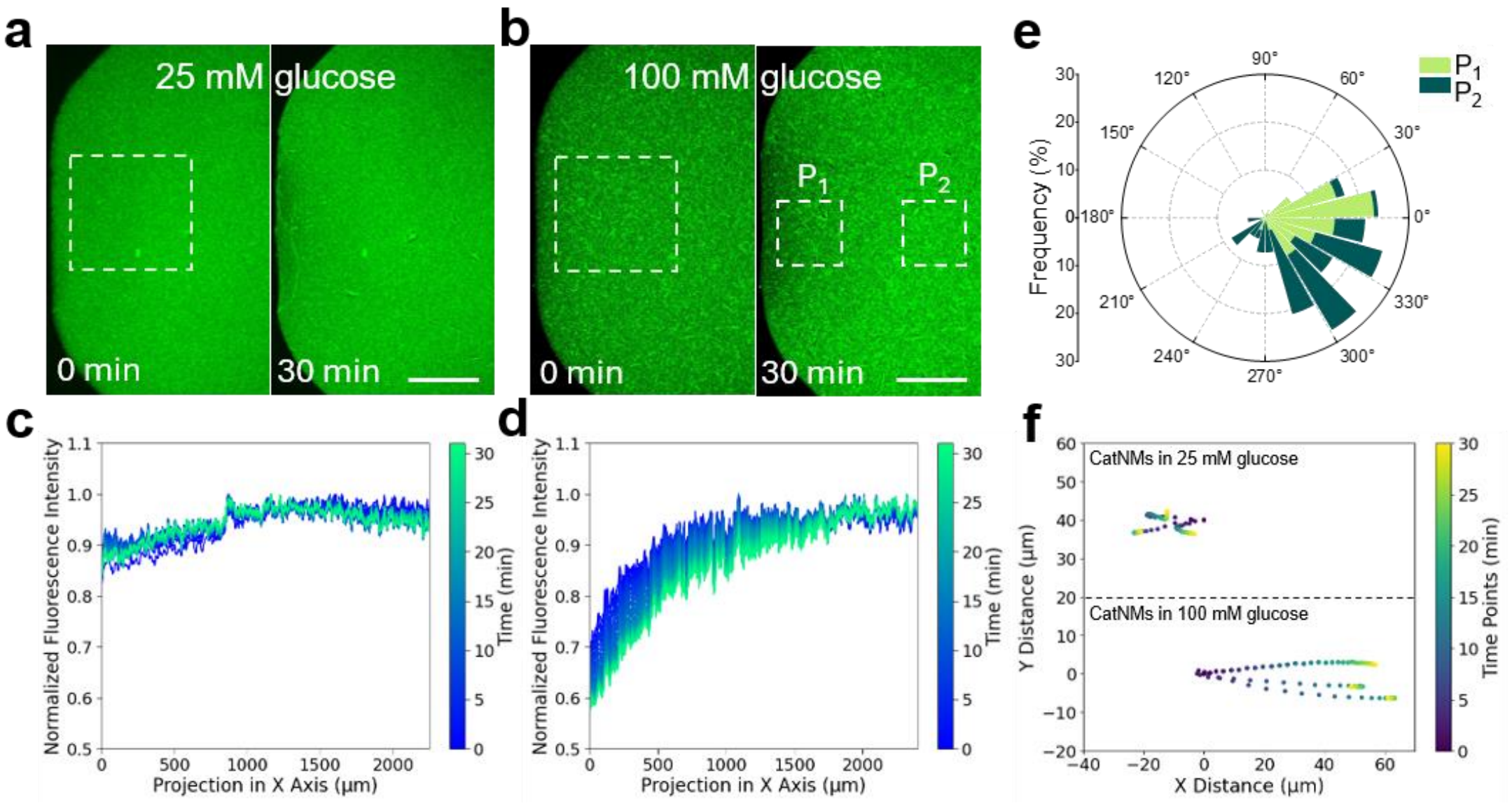

**Extended Fig. E2 CatNMs repelled by glucose gradients.** Fluorescence images of CatNMs distribution under glucose gradients at **a** 25 mM (in PBS 0.1× solutions) and **b** 100 mM (in PBS 0.1× solutions), captured at 0 min and 30 min. Scale bar: 1 mm. **c-d** Normalized fluorescence intensity profiles for the ROIs (dashed squares) in **a-b**. **e** Polar map of ENM trajectory directionality after 5 s in regions $P_1$ and $P_2$ (N=100). **f** Time-dependent displacement of the fluorescence centroid in the ROIs (squares in **a-b**) across conditions. N = 3 independent experiments.

# Supporting Information

## Controlled Chemical Signaling between Enzymatic Nanomotors

*Shuqin Chen[1,2], Giorgio Lovato[3], Oriol Jutglar Soler[1], Daniel Sánchez-deAlcázar[1], Ramin Golestanian[3,4*], Samuel Sánchez[1,5*]*

[1]*Institute for Bioengineering of Catalonia (IBEC), The Barcelona Institute for Science and Technology (BIST), Baldiri i Reixac 10-12, Barcelona 08028, Spain*

[2]*Department of Physics, University of Barcelona, Martí i Franquès 1, Barcelona 08028, Spain*

[3]*Max Planck Institute for Dynamics and Self-Organization (MPI-DS), D-37077 Göttingen, Germany*

[4]*Rudolf Peierls Centre for Theoretical Physics, University of Oxford, Oxford OX1 3PU, United Kingdom*

[5]*Catalan Institute for Research and Advanced Studies (ICREA), Passeig Lluís Companys 23, Barcelona 08010, Spain*

Corresponding authors:

*Ramin Golestanian: ramin.golestanian@ds.mpg.de

*Samuel Sánchez: ssanchez@ibecbarcelona.eu

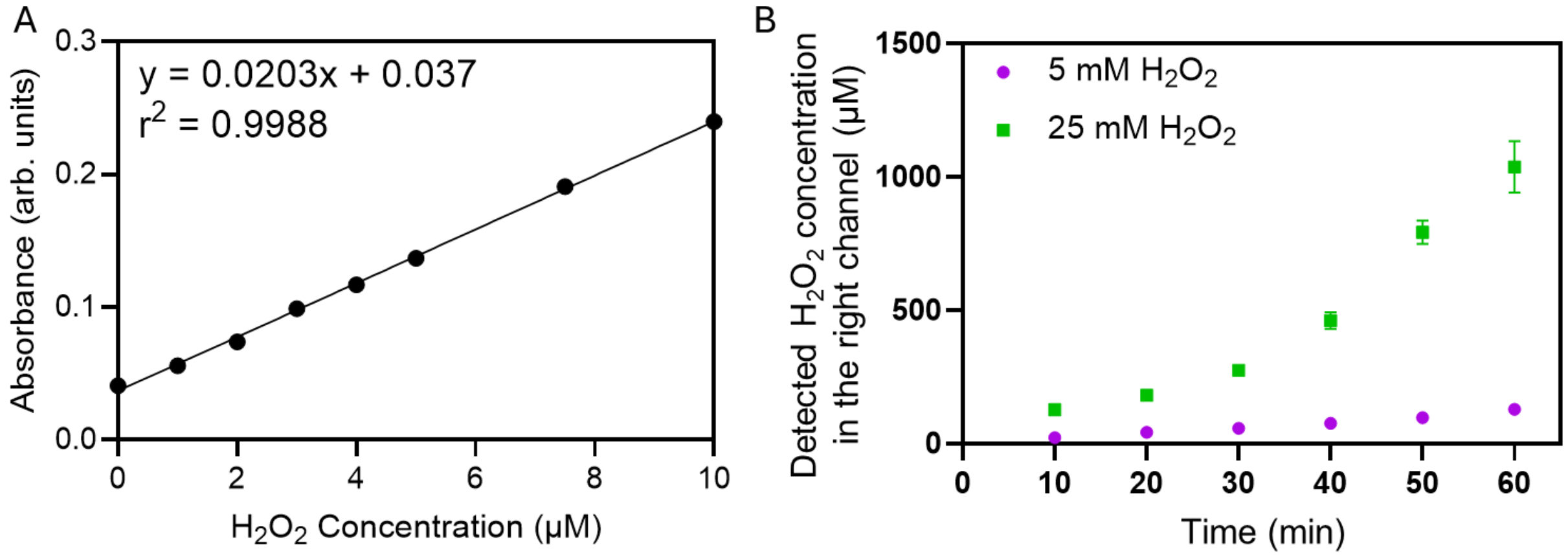

**Fig. S1 Detection of $H_2O_2$ diffusion.** (A) Standard curve of $H_2O_2$ concentration based on the colorimetric response of Amplex Red. (B) $H_2O_2$ (5 mM or 25 mM) diffuses from the source toward the right channel, with $H_2O_2$ concentration in the right channel measured every 10 min. N=3 independent experiments.

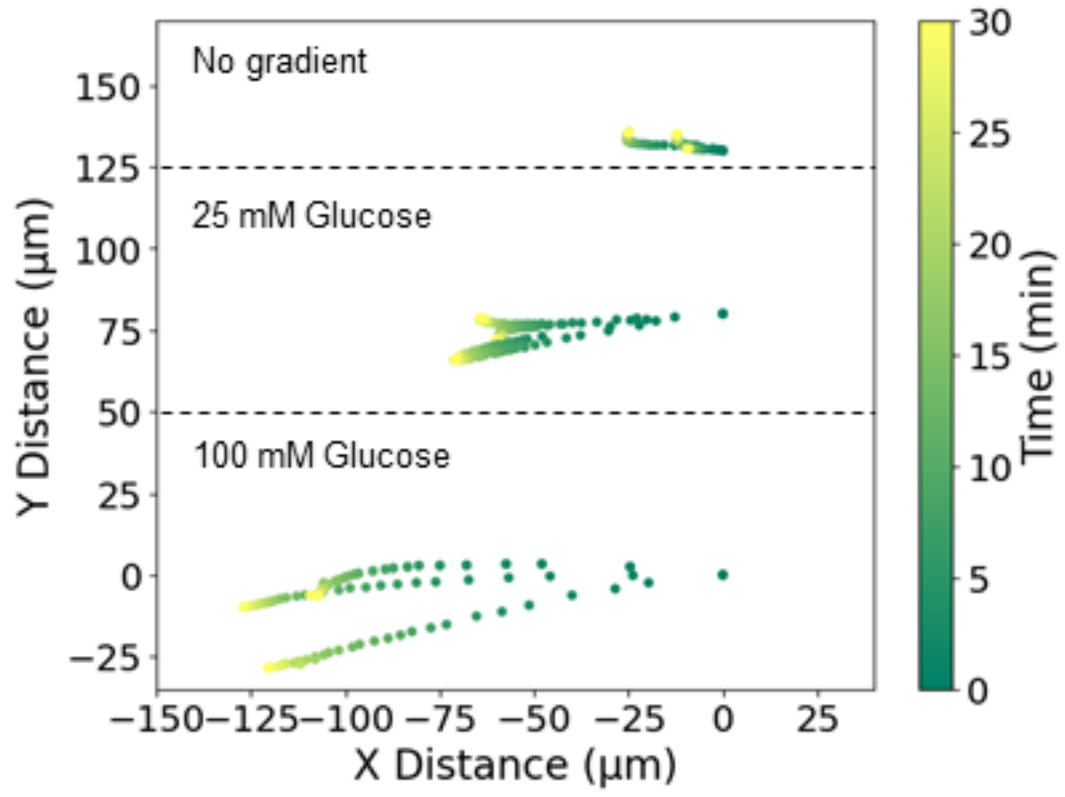

**Fig. S2 Time-dependent displacement of the fluorescence centroid of GOxNMs across different conditions**. The horizontal displacement increases with the glucose fuel concentration. N = 3 independent experiments.

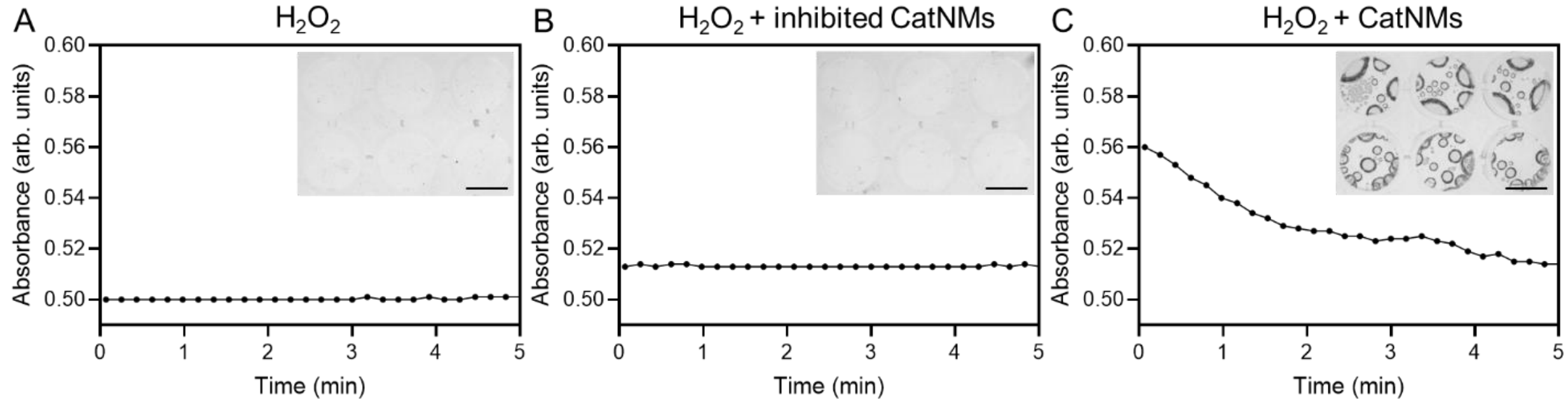

**Fig. S3 Assessment of nanomotor activity.** Absorbance of $H_2O_2$ at 240 nm under different conditions: (A) $H_2O_2$, (B) $H_2O_2$ + activity-inhibited CatNMs, and (C) $H_2O_2$ + active CatNMs. The insets show images of $H_2O_2$ catalyzation after nanomotor addition for 30 min. Scale bars: 5 mm.

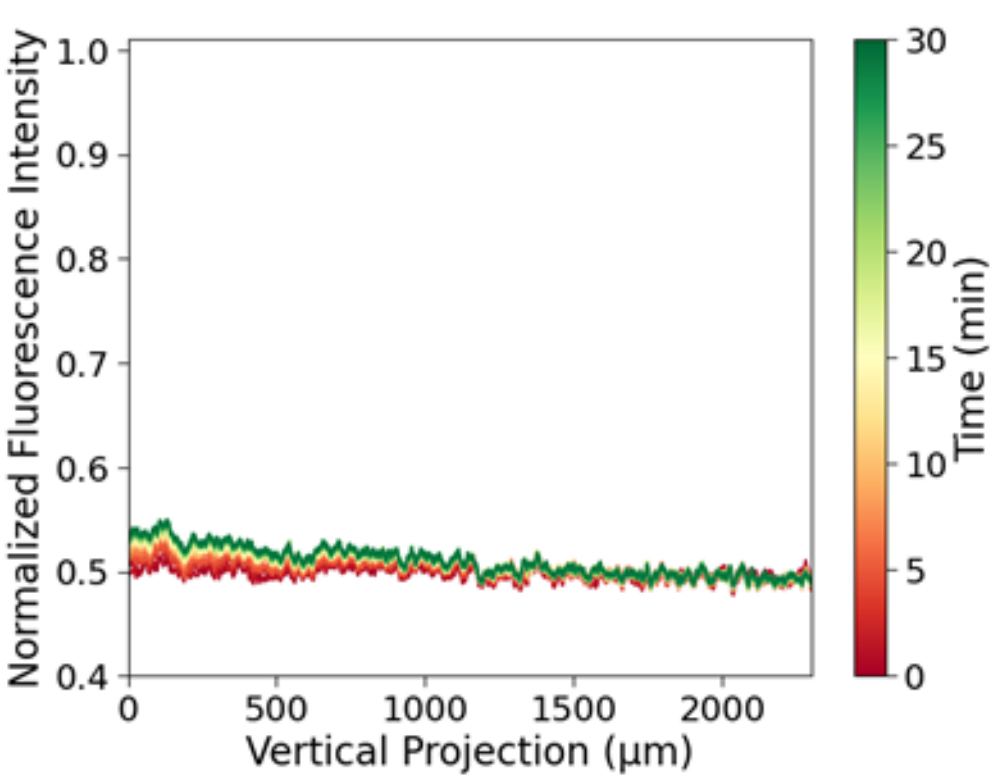

**Fig. S4 Normalized fluorescence intensity profiles of activity-inhibited CatNMs in 25 mM $H_2O_2$ gradients.** The region of interest for analysis was selected from the dashed squares in Figure 3**n**.

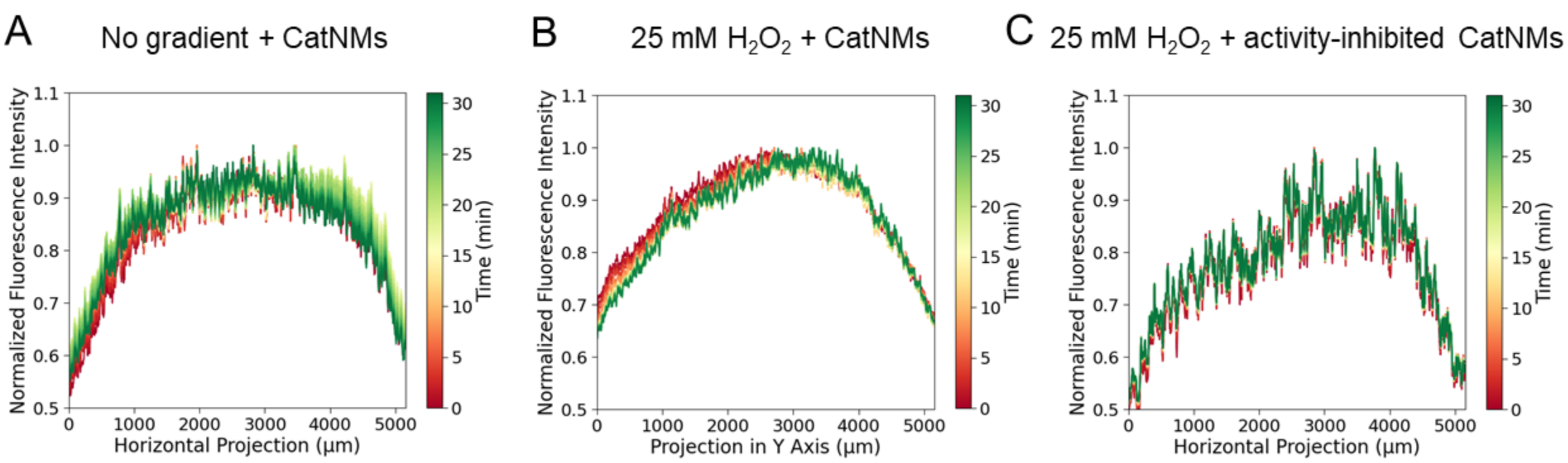

**Fig. S5 Fluorescence intensity profiles quantified by horizontal projection**. A. CatNMs under no gradient conditions. B. CatNMs in a 25 mM $H_2O_2$ gradient. C. Activity-inhibited CatNMs in a 25 mM $H_2O_2$ gradient.

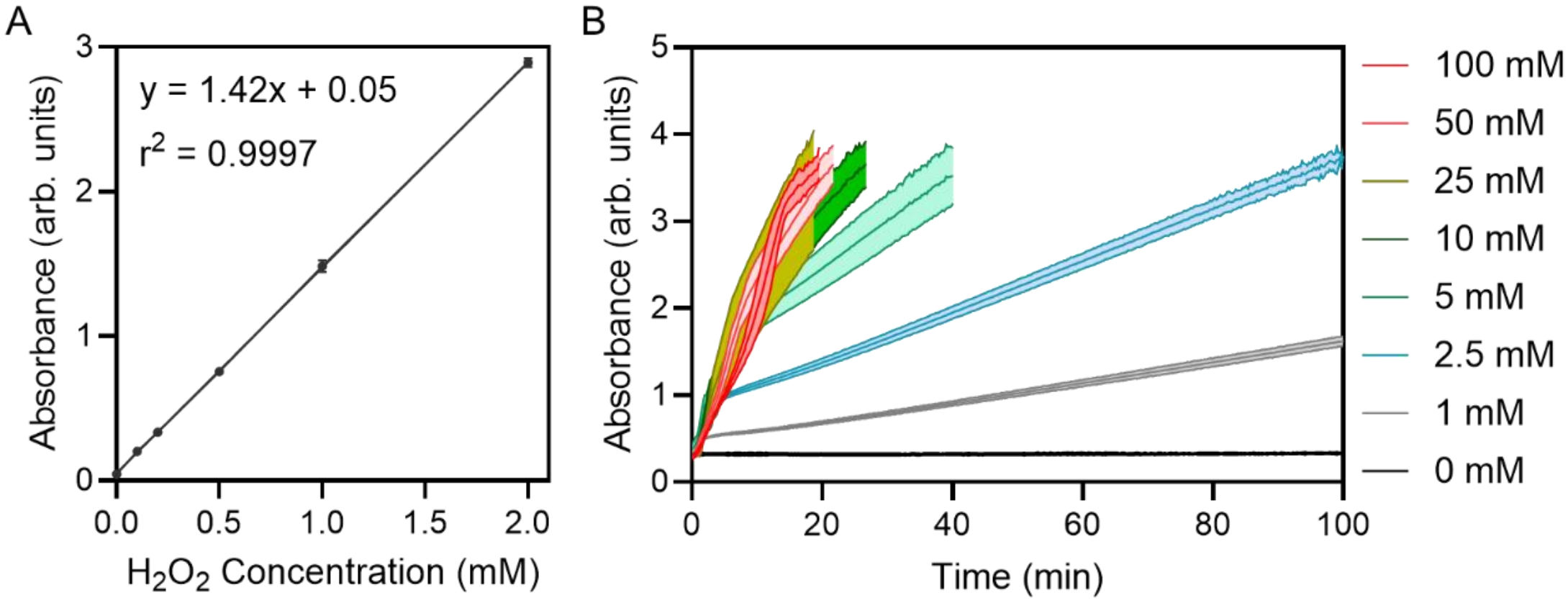

**Fig. S6 Enzymatic activity of GOxNMs under varying glucose concentrations.** A. Standard $H_2O_2$ calibration curve quantified through ABTS/HRP-based colorimetric methods. B. The absorbance spectrum corresponds to the oxidation of ABTS in a coupled reaction with HRP and GOxNMs. The absorbance plateaus within the detection interval because the color of the products exceeds the spectrophotometer's linear detection range.

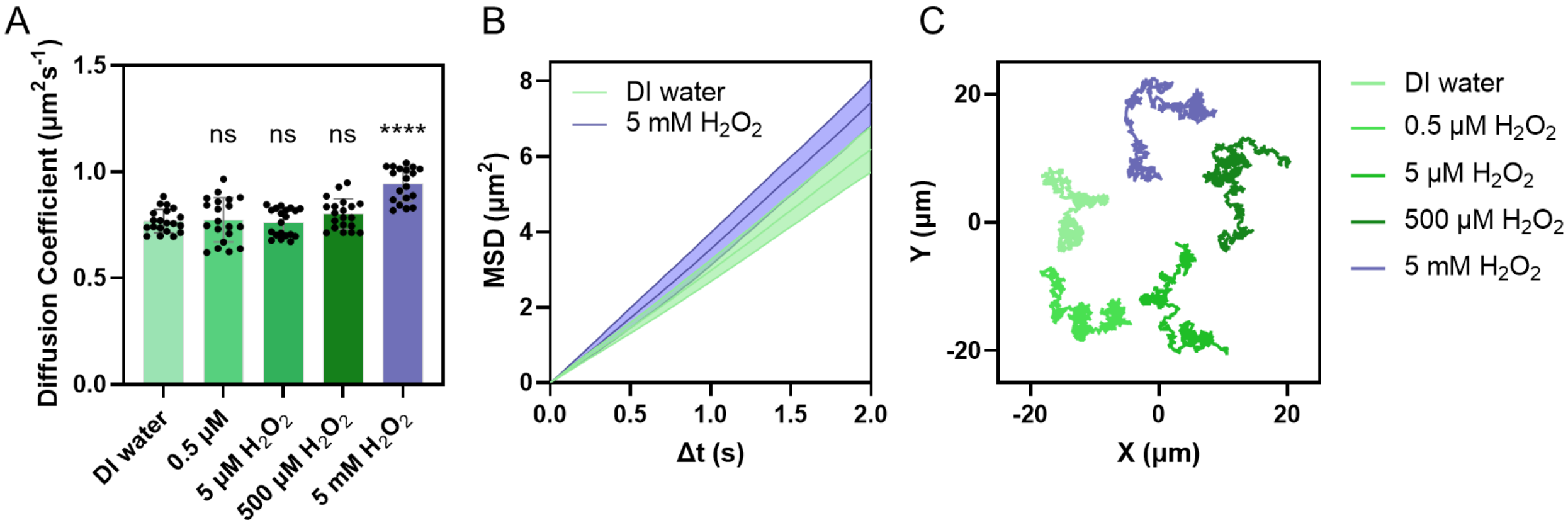

**Fig. S7 Motion analysis of single CatNMs at low $H_2O_2$ concentrations.** A. Diffusion coefficient of CatNMs in 0-5 mM $H_2O_2$ solutions. Significant difference is analyzed by student's t-test: ****$P < 0.0001$; ns = not significant ($P > 0.05$). N = 20. B. Mean square displacement of CatNMs in DI water and in 5 mM $H_2O_2$. C. Representative trajectories of CatNMs in different concentrations of $H_2O_2$.

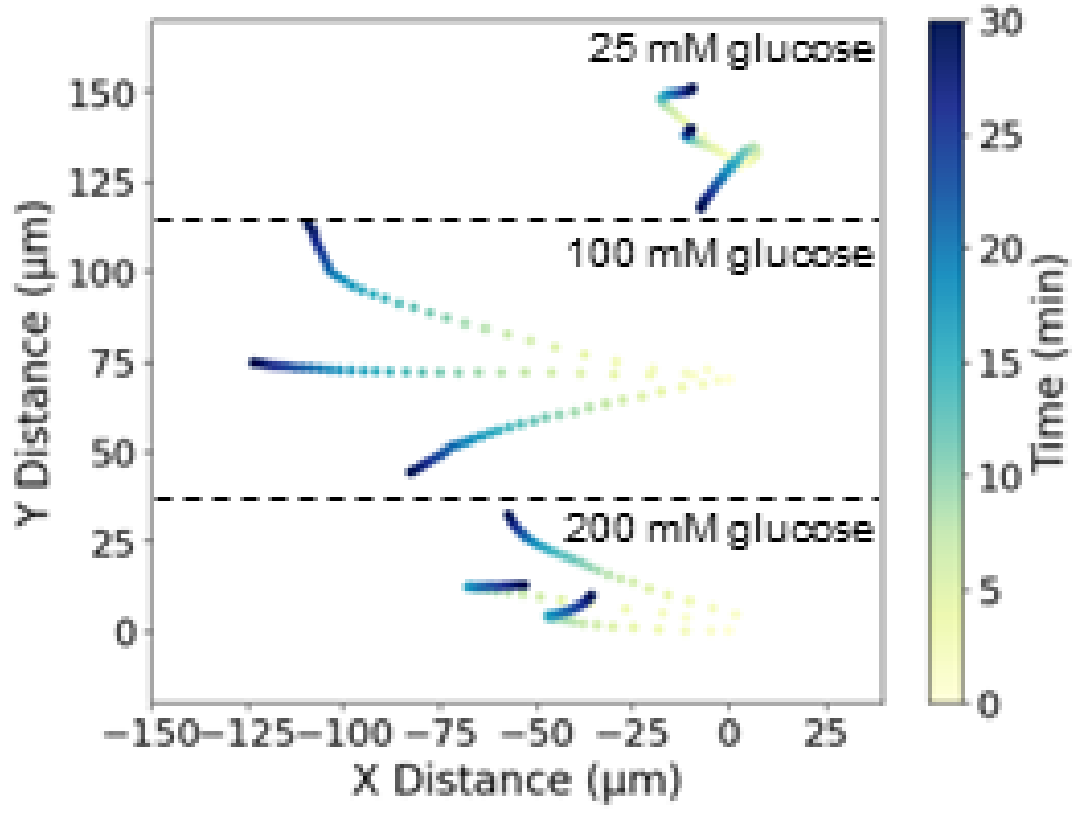

**Fig. S8 Time-dependent displacement of the fluorescence centroid of CatNMs across different conditions**. The horizontal displacement increases with the glucose fuel concentration. N = 3 independent experiments.

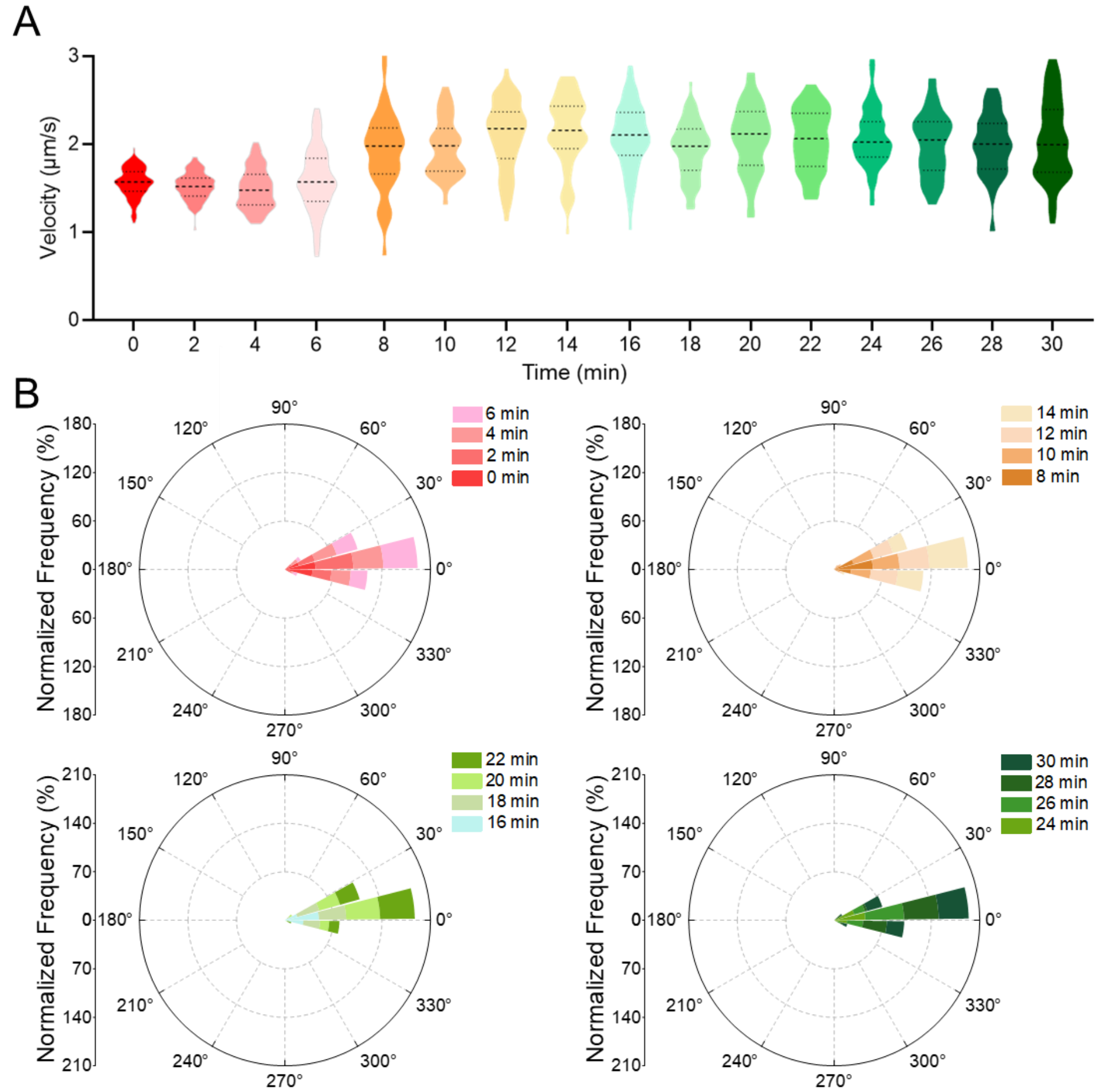

**Fig. S9 Motion analysis of CatNMs (in near region) in self-generated $H_2O_2$ gradients by 200 mM glucose diffusing through GOxNMs-embedded gel.** (A) Velocity analysis for nanomotors in position near the fuel within 50 min started from the introduction of fuel. CatNMs near the gel initially exhibit a high drifting velocity (1.6 μm/s), which slow within 5 minutes before recovering to a similarly high value. (B) Polar map of nanomotor orientations within 0-50 min. These nanomotors display initial motion toward the gel, followed by gradual repulsion due to the chemical gradient.

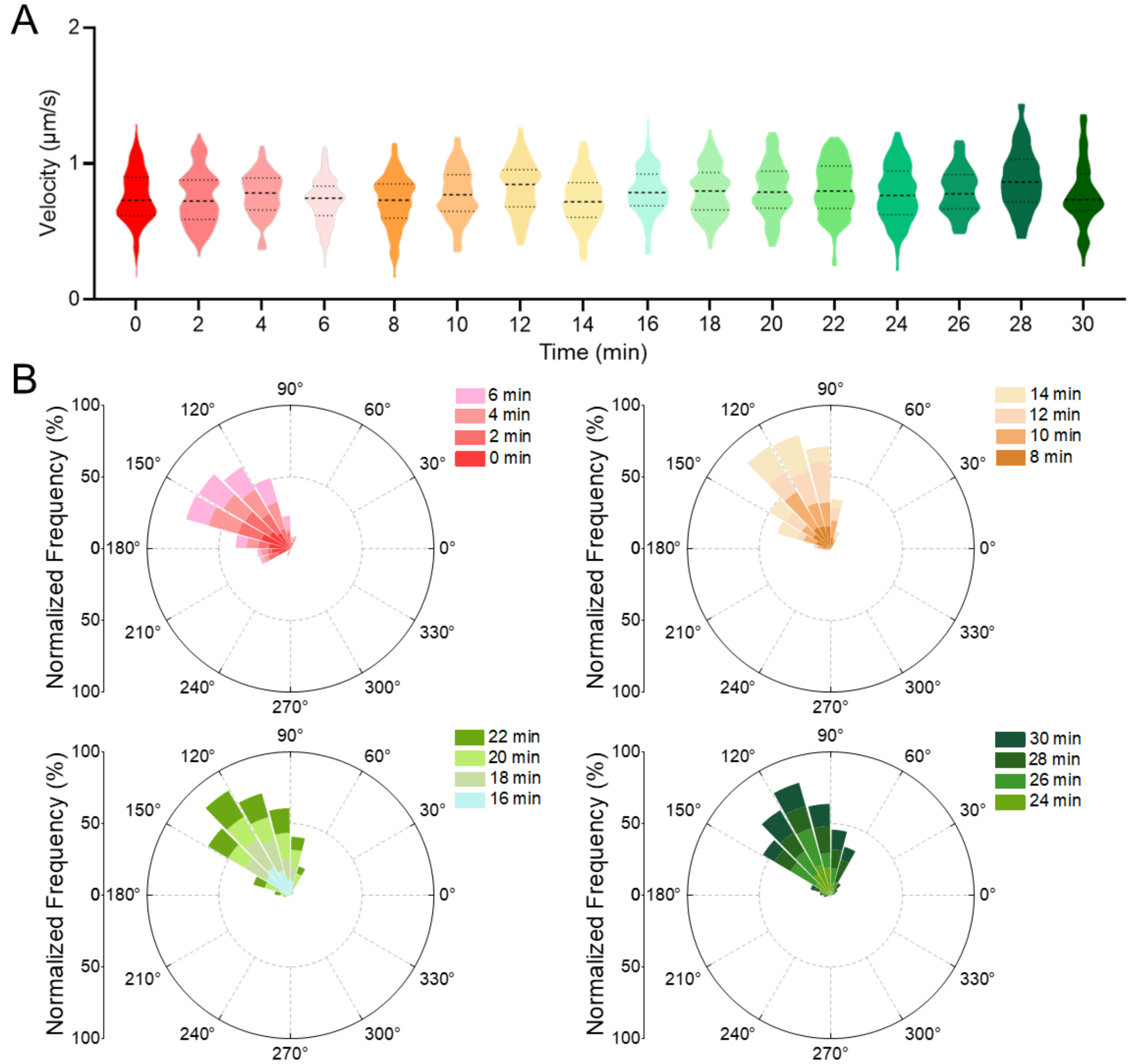

**Fig. S10 Motion analysis of CatNMs (in far region) in self-generated $H_2O_2$ gradients by 200 mM glucose diffusing through GOxNMs-embedded gel.** (A) Velocity analysis for particles in position far from the fuel within 50 min started from the introduction of fuel. CatNMs farther from the gel moved at a constant 0.8 μm/s, primarily toward the gel side. (B) Polar map of nanomotor orientations within 0-50 min.

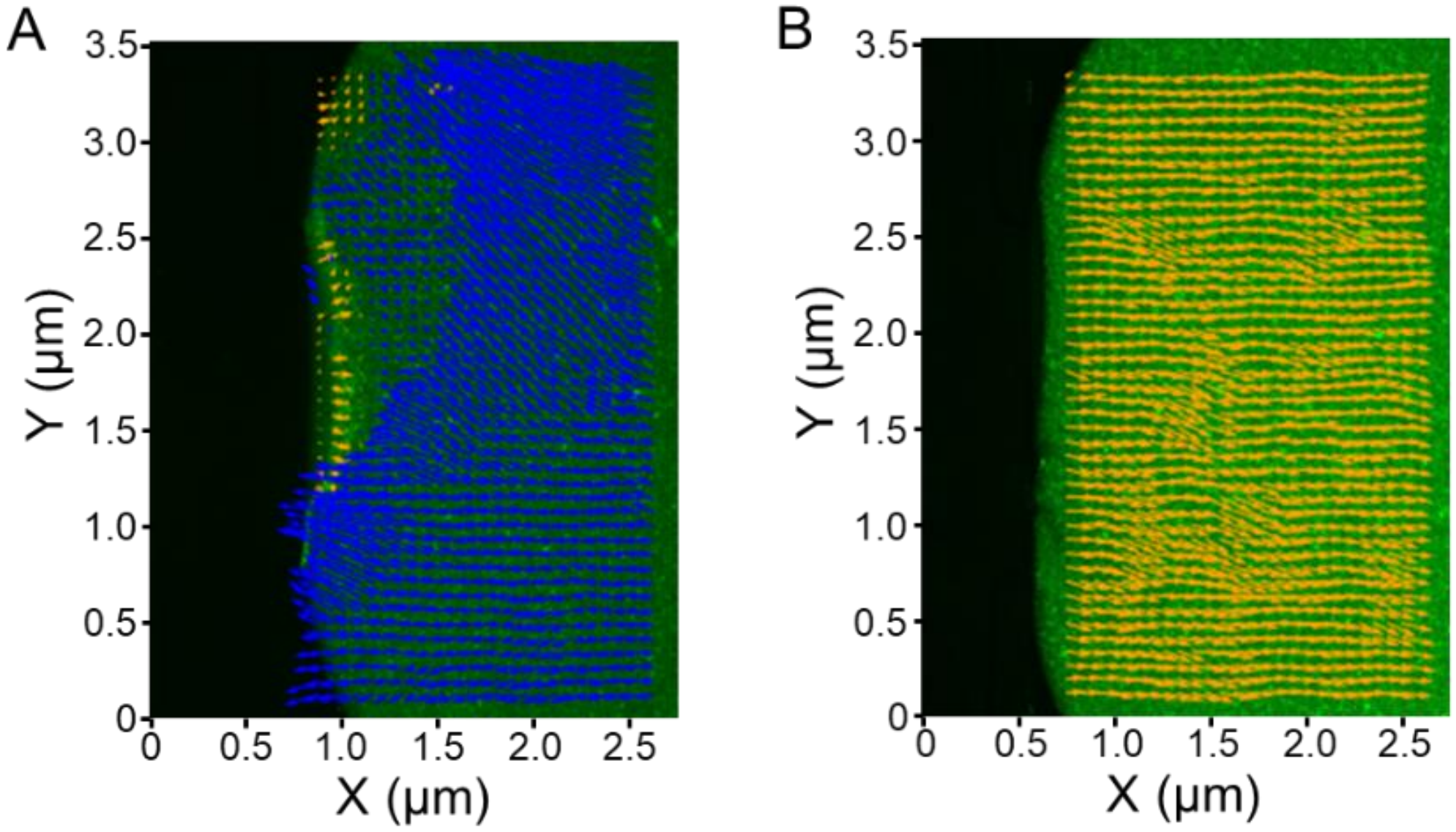

**Fig. S11 Particle image velocimetry analysis of CatNMs' collective motion.** (A) When the central gel was prepared with deionized water and exposed to a gradient of $H_2O_2$, the CatNMs collectively migrated toward the fuel source (blue arrows), whereas an exclusion zone near the gel side repelled them (yellow arrows). (B) Preparing the central gel with 0.1× PBS buffer solution effectively prevented the formation of an exclusion zone. However, under a glucose gradient (100 mM), the absence of PBS buffer in the CatNMs will lead to osmotic flow and induced negative collective migration. All CatNMs moved away from the gel side (orange arrows).

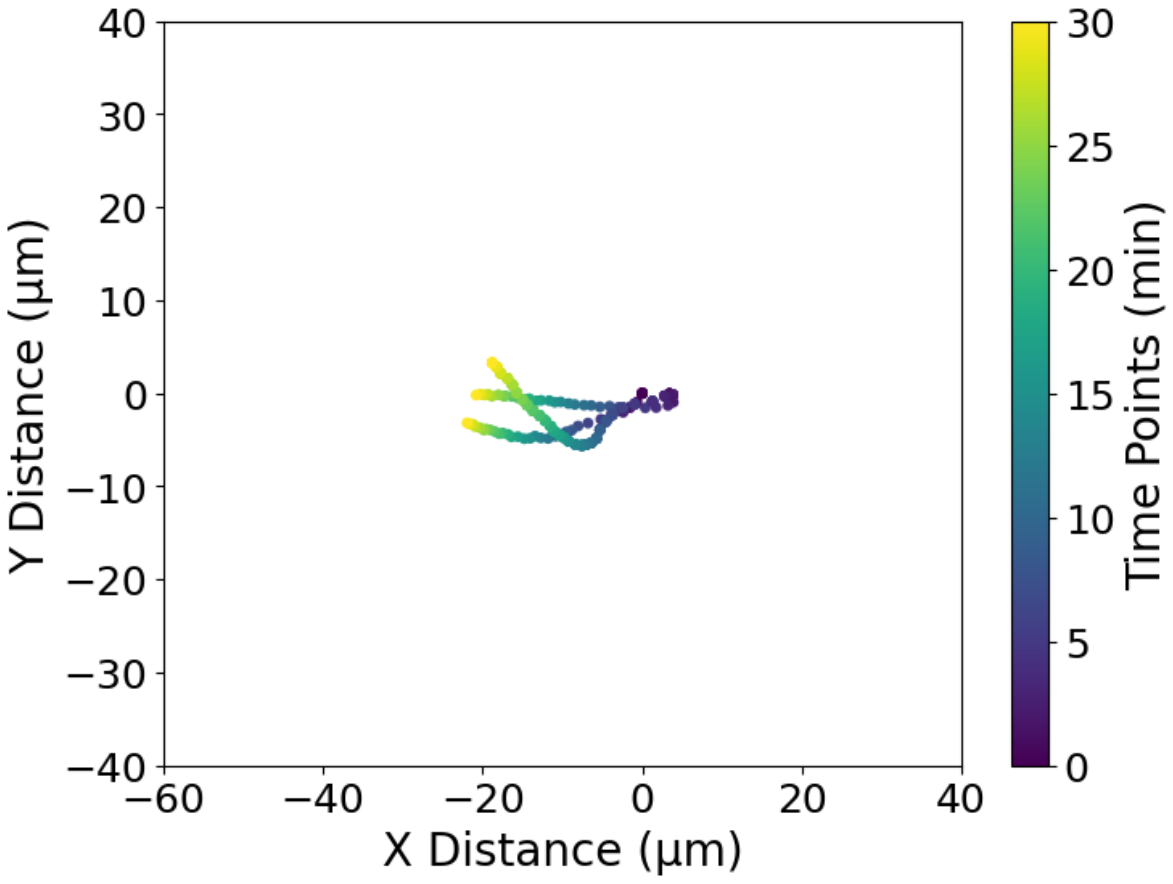

**Fig. S12 Time-dependent displacement of the fluorescence centroid.** CatNMs in 300 mM urea solutions. N = 3 experiments.

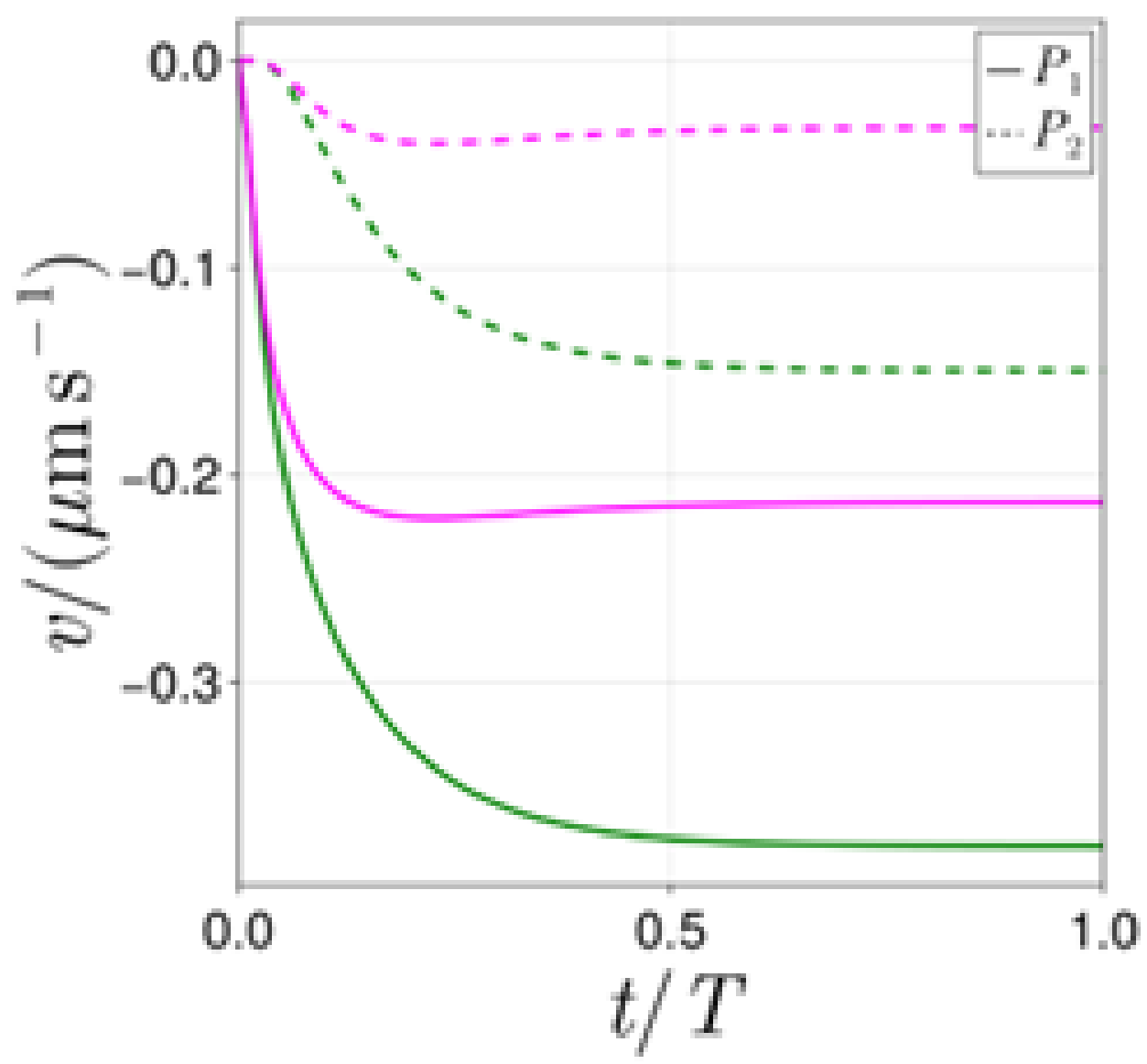

**Fig. S13 Time series of the velocities at the two positions for the combinations indicated by the same color dots in Fig. 6c.** At around 1/4 of the simulation time, the velocities relax towards a plateau value. In the sublinear class, the two velocities differ a lot, while in the superliner class, they are closer in magnitude.

**Table S1. The amount of linked glucose oxidase and catalase by BCA analysis**

| | Initial enzyme added (µg/mL) | | | Supernatant 1 (µg/mL) | | | Supernatant 2 (µg/mL) | | | Linked enzyme (µg/mL) |
|---|---|---|---|---|---|---|---|---|---|---|
| GOxNMs | 317.7 | 296.3 | 260.0 | 198.6 | 191.2 | 190.2 | 0 | 0 | 0 | 98.0 |
| Catalase | 372.6 | 361.4 | 366.0 | 265.6 | 256.3 | 257.2 | 0 | 0 | 0 | 107.0 |